\documentclass[fleqn,usenatbib,onecolumn]{mnras}

\usepackage{newtxtext,newtxmath}
\usepackage[T1]{fontenc}

\DeclareRobustCommand{\VAN}[3]{#2}
\let\VANthebibliography\thebibliography
\def\thebibliography{\DeclareRobustCommand{\VAN}[3]{##3}\VANthebibliography}

\usepackage{graphicx}	
\usepackage{amsmath}	
\usepackage{amssymb}	
\usepackage{wasysym}           
\usepackage{graphicx}
\usepackage{epstopdf}
\usepackage{mathrsfs}
\usepackage{anyfontsize}
\usepackage{natbib}
\usepackage{color}
\usepackage{lipsum}
\usepackage{diagbox}

\title[Structured, relativistic jets]{Structured, relativistic jets driven by radiation}

\author[Coughlin \& Begelman]{
Eric R.~Coughlin,$^{1, 2}$\thanks{E-mail: ecoughli@syr.edu}
and Mitchell C.~Begelman$^{3,4}$
\\
$^{1}$Department of Physics, Syracuse University, Syracuse, NY 13244, USA \\
$^{2}$Department of Astrophysical Sciences, Princeton University, Princeton, NJ 08544, USA\\
$^{3}$JILA, University of Colorado and NIST, 440 UCB, Boulder, CO 80309, USA\\
$^{4}$Department of Astrophysical and Planetary Sciences, 391 UCB, Boulder, CO 80309, USA
}

\date{Accepted XXX. Received YYY; in original form ZZZ}

\pubyear{2020}

\begin{document}
\label{firstpage}
\pagerange{\pageref{firstpage}--\pageref{lastpage}}
\maketitle

\begin{abstract}
Relativistic jets, or highly collimated and fast-moving outflows, are endemic to many astrophysical phenomena. The jets produced by gamma-ray bursts and tidal disruption events are accompanied by the accretion of material onto a black hole or neutron star, with the accretion rate exceeding the Eddington limit of the compact object by orders of magnitude. In such systems, radiation dominates the energy-momentum budget of the outflow, and the dynamical evolution of the jet is governed by the equations of radiation hydrodynamics. Here we show that there are analytic solutions to the equations of radiation hydrodynamics in the viscous (i.e., diffusive) regime that describe structured, relativistic jets, which consist of a fast-moving, highly relativistic core surrounded by a slower-moving, less relativistic sheath. In these solutions, the slower-moving, outer sheath contains most of the mass, and the jet structure is mediated by local anisotropies in the radiation field. We show that, depending on the pressure and density profile of the ambient medium, the angular profile of the jet Lorentz factor is Gaussian or falls off even more steeply with angle. These solutions have implications for the nature of jet production and evolution in hyperaccreting systems, and demonstrate that such jets -- and the corresponding jet structure -- can be sustained entirely by radiative processes. We discuss the implications of these findings in the context of jetted tidal disruption events and short and long gamma-ray bursts.
\end{abstract}

\begin{keywords}
black hole physics --- hydrodynamics --- methods: analytical --- radiation: dynamics --- radiative processes --- relativistic processes
\end{keywords}

\section{Introduction}
Astrophysical explosions are initiated by the injection of a large amount of energy in a variety of forms. For hyperaccreting black holes, the energy is in the form of an intense radiation field that is generated by viscous dissipation as matter accretes. As the radiation thermalizes with and couples to the gas, the material gains momentum from anisotropies in the radiation field and an outflow is generated. Such radiation-driven outflows from black holes are likely formed in tidal disruption events (TDEs; e.g., \citealt{strubbe09, coughlin14b, metzger16}) and gamma-ray bursts (GRBs; e.g., \citealt{meszaros92, rees92}), where the accretion rate onto the black hole can be many orders of magnitude above the Eddington limit. 

The pressure gradient established by the radiation field can be sufficient, in these systems, to drive an optically thick and roughly spherically symmetric outflow. However, GRBs (and to a lesser extent TDEs; e.g., \citealt{bloom11, burrows11, levan11, zauderer11, cenko12, brown15}) are often characterized by the existence of a highly collimated, relativistic jet of faster-moving material, inferred from the detection of a ``break'' in the emission properties (e.g., \citealt{rhoads99, sari99, harrison99, panaitescu02, berger14}). Moreover, the observations of the afterglow of the neutron star-neutron star merger GW/GRB 170817 \citep{abbott17} have found that these jets possess an inherent degree of \emph{structure}, and exhibit angular variations in the jet Lorentz factor (e.g., \citealt{lamb17, lyman18, fong19, troja19}). 

The origin of the collimation and launching of the jets from these systems could be intrinsically magnetic \citep{blandford77}, and additional collimation and structure formation will be provided by the ambient medium in the immediate vicinity of the launching mechanism \citep{bromberg07, kohler12, kath19}.  To the extent that these are the only agents that provide the source of structure within the outflow, the nature of the jet structure (e.g., the precise velocity profile as a function of polar angle from the jet axis) is determined exclusively by conditions that are extrinsic to the jet itself and depend sensitively on the magnetization of the gas. 

Another possibility is that the jet structure is established by local, viscous-like interactions within the fluid that comprises the jet itself. When the jet is highly magnetized, the viscous dissipation and coupling is generated by the wave-like interactions between the magnetic field and the gas (i.e., a Braginskii-like viscosity; e.g., \citealt{braginskii65, chandra15, finazzo16, foucart16}). The energetics of the jets from hyperaccreting black holes should, however, be dominated by \emph{radiation}, or photons created during the accretion process, even if the initial launching mechanism is magnetic. In this scenario, the source of the viscosity is in the form of local, small-scale (i.e., over the photon mean free path, which in these optically thick systems is typically much smaller than the scale of the jet itself but much longer than the interparticle mean free path; e.g., \citealt{arav92}) anisotropies in the radiation field, the anisotropies themselves created by relativistic Doppler beaming as perceived in the rest frame of a given fluid element. The shear across a relativistic, structured jet can therefore be established in situ and maintained by the same source of energy that powers the outflow\footnote{Here we consider the case where the flow is laminar; see \citet{zrake19} for an analysis of the effects of radiation viscosity in the presence of turbulence.}. 

Here we show that structured jets naturally arise from the viscous-like interactions between the gas and photons within a relativistic outflow. In Section \ref{sec:basic} we present the equations of radiation hydrodynamics in the viscous limit, which govern the dynamical evolution of the fluid in the radiation-dominated, optically-thick regime, which is appropriate for hyperaccreting systems. In Section \ref{sec:jets} we derive self-similar solutions that describe a relativistic jet {}{powered by a given energy injection rate} and embedded in an ambient medium, and we show that the fluid density and jet Lorentz factor display substantial variation (i.e., structure) {}{over the angular width of the jet}. {}{We show that, when the jet is ultra-relativistic, the self-similar equations possess analytic and closed-form solutions, with the jet Lorentz factor falling off as a Gaussian in angle from the jet axis. The implications of these solutions are discussed in the context of jetted tidal disruption events and gamma-ray bursts in Section \ref{sec:implications}}. We summarize and offer concluding remarks in Section \ref{sec:summary}.

\section{Equations and Basic Assumptions}
\label{sec:basic}
The equations of radiation hydrodynamics describe an optically thick outflow in which matter and radiation dominate the energy-momentum budget of the fluid. In tensor notation, these equations are written (e.g., \citealt{mihalas84})

\begin{equation}
\nabla_{\mu}\left[T^{\mu\nu}+R^{\mu\nu}\right] = 0, \label{radhydro}
\end{equation}
where

\begin{equation}
T^{\mu\nu} = \rho'U^{\mu}U^{\nu}
\end{equation}
is the energy-momentum tensor of the matter with comoving density $\rho'$ and four-velocity $U^{\mu}$, and we adopted the Einstein summation convention such that repeated upper and lower indices imply summation over that repeated index. $R^{\mu\nu}$ is the energy-momentum tensor of the radiation field. In the limit that the fluid is infinitely optically thick in the comoving frame, so that the photon mean free path is zero and the radiation field is exactly isotropic (again, within that frame), this tensor is given by

\begin{equation}
R^{\mu\nu} = R^{\mu\nu}_{\rm iso} = 3p'U^{\mu}U^{\nu}+p'\Pi^{\mu\nu}, \label{Rmunuiso}
\end{equation}
where $p'$ is the comoving radiation pressure, which itself can be related to integrals of the photon distribution function over the photon four-momentum, and $\Pi^{\mu\nu} = U^{\mu}U^{\nu}+g^{\mu\nu}$ is the projection tensor with $g^{\mu\nu}$ the inverse of the metric. Here, and for the remainder of this paper {}{(except in a few situations where it is convenient to reintroduce it explicitly)}, we use units in which the speed of light is equal to one. 

While Equation \eqref{Rmunuiso} is a good approximation for the radiation field in optically thick media, it ignores finite mean free path effects. When such terms are maintained, in the comoving frame of a given fluid element scatterers ``see'' small anisotropies in the radiation field that are induced by pressure gradients (analogous to conduction) and variation in the fluid velocity (analogous to viscosity). Keeping first-order terms in the photon mean free path results in a photon energy-momentum tensor of the form $R^{\mu\nu} = R^{\mu\nu}_{\rm iso}+R^{\mu\nu}_{\rm vis}$, where (for photon-matter interactions via Thomson scattering in the comoving frame of a fluid element)

\begin{multline}
R^{\mu\nu}_{\rm vis} = -\frac{40}{9}\frac{p'}{\rho'\kappa}\left(U^{\mu}U^{\nu}+\frac{1}{3}\Pi^{\mu\nu}\right)\nabla_{\alpha}U^{\alpha}
-\frac{8}{9}\frac{p'}{\rho'\kappa}\Pi^{\mu\sigma}\Pi^{\nu\beta}\left(\nabla_{\sigma}U_{\beta}+\nabla_{\beta}U_{\sigma}-\frac{2}{3}g_{\sigma\beta}\nabla_{\alpha}U^{\alpha}\right) \\
-\frac{1}{\rho'\kappa}\left(U^{\mu}\Pi^{\nu\sigma}+U^{\nu}\Pi^{\mu\sigma}\right)\left(4p'U^{\beta}\nabla_{\beta}U_{\sigma}+\nabla_{\sigma}p'\right). \label{Rmunuvis}
\end{multline}
Here $\kappa$ is the opacity of the scatterers within the fluid, and the various numerical factors arise from the aspherical nature of the Thomson cross section of the electrons in the comoving frame. This equation has the general form of a relativistic, viscous energy-momentum tensor as proposed by \citet{eckart40}, and each one of these terms has a physical interpretation; we refer the reader to \citet{coughlin14} for a more in-depth discussion of the meaning of each of these terms (see also \citealt{weinberg71, loeb92}). 

Here we assume that the production of an intense radiation field through accretion establishes a large influx of energy near the compact object, which accelerates the fluid. The existence of an accretion flow establishes a non-spherical distribution of mass around the black hole, which anisotropically funnels radiation (and energy and momentum) along the rotational axis of that flow. Owing to this enhanced energy and momentum flux, we expect the outflow velocity along the axis to be correspondingly boosted, and this physical setup therefore naturally establishes a structured jet: a fast-moving core surrounded by a slower, less-collimated outflow\footnote{{}{We will also assume that as material accretes it loses enough angular momentum such that the rotational velocity of the outflow is much less than the radial component directed away from the black hole. It may be possible to generalize the solutions discussed here by including an additional, centrifugal term, but we do not pursue this route any further.}}. Following the establishment of the initial anisotropy of the outflow and the radiation field, small-scale (i.e., on the scale of the photon mean free path) variations in the radiation field -- and the corresponding viscous interaction between the photons and the fluid -- provide the means to locally mediate the shear within the outflow and establish the jet structure. 

\begin{figure}
   \centering
   \includegraphics[width=0.795\textwidth]{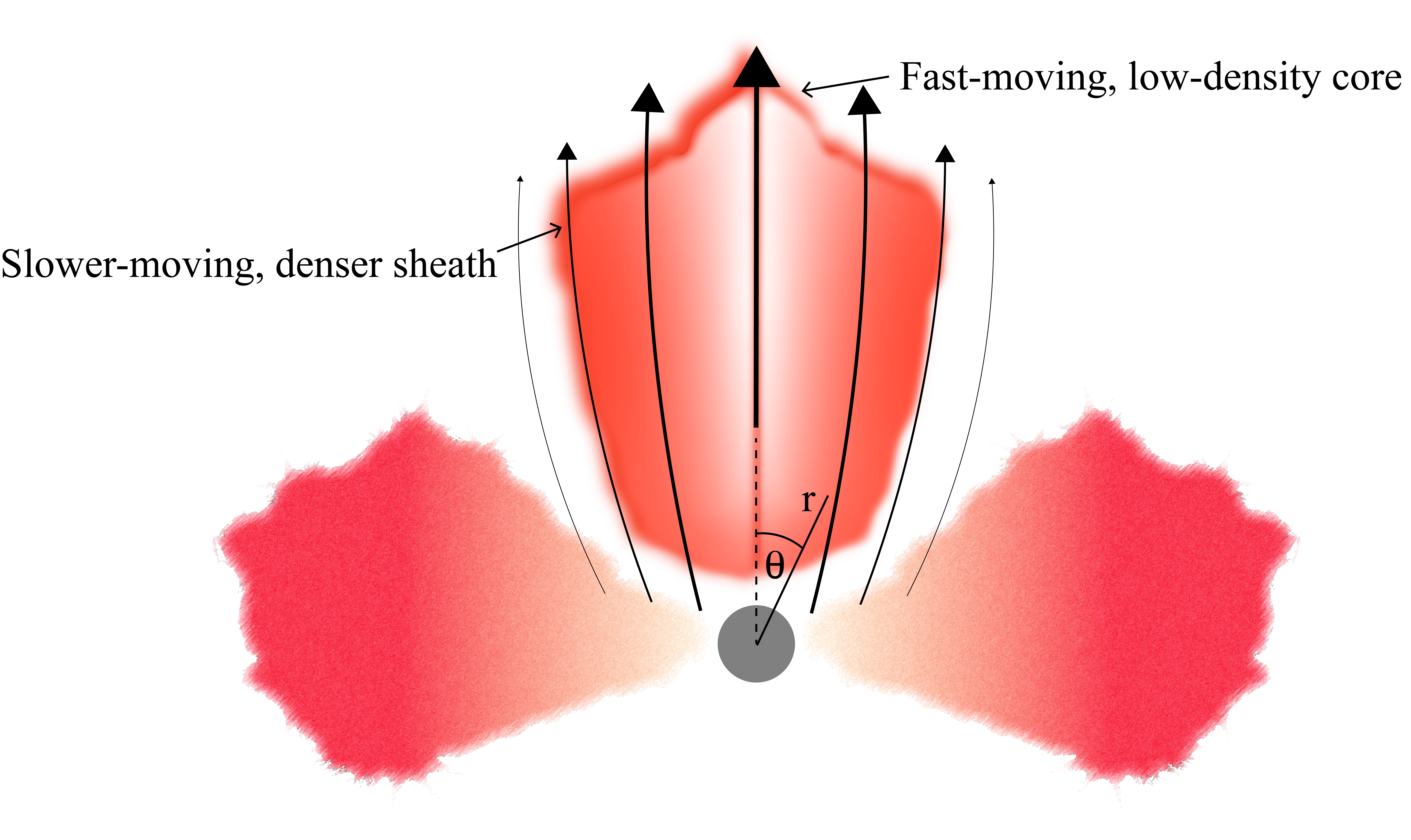} 
   \caption{{}{A schematic of a structured jet: the hyperaccreting black hole-disc system (shown at the bottom) establishes a large injection of energy that accelerates material in the form of a jet. The fastest-moving, relatively low-density material is along the axis of the jet, which we define as the core, while the slower-moving sheath -- at larger $\theta$, where $\theta$ is the angle measured relative to the jet axis -- contains most of the mass. The curves indicate qualitatively the streamlines, with arrows giving the direction of the flow and larger arrows/thicker lines corresponding to greater Lorentz factors. The fact that the streamlines curve toward the jet axis at large $\theta$ indicates that the jet is collimated by the ambient medium, and that the jet entrains mass from its surroundings. The overall angular width over which the flow assumes this form is on the order of the boundary layer thickness $\delta\theta \sim 1/\sqrt{\rho'\kappa\Gamma r}$ (see the text for further discussion).}}
   \label{fig:black_hole_jet}
\end{figure}

{}{Figure \ref{fig:black_hole_jet} illustrates the qualitative picture of such a structured jet and the geometry; at large angles $\theta$ (where the angle $\theta$ is measured from the axis of the jet) and radii $r$, the outflow is in the form of a slower-moving, denser sheath. Along the axis of the jet the material is more relativistic and possesses a smaller density of scatterers, and the shear between the slow-moving sheath and fast-moving core is mediated by radiation. The curves give a qualitative depiction of the streamlines, with the arrowheads indicating the direction of the flow; the width of the line gives a measure of the magnitude of the four-velocity. The fact that the streamlines curve toward the axis indicates that the jet is collimated by the ambient medium and is parasitic, drawing in scatterers from the surrounding sheath. The velocity along the axis is governed by a competition between the increase in the inertia contained in the outflow, which decelerates the flow and arises both from the lateral expansion of the jet and entrainment from the sheath, and the pressure gradient of the ambient gas, which acts as an accelerating agent. The conditions at the base of the outflow, including the initial Lorentz factor, are established by the injection of energy near the hyperaccreting object.}

Because the radiation field dominates the energy-momentum budget of the fluid, the transition from the fast-moving core to the slower moving sheath is mediated by the photons within the flow. When the fluid is optically thick, photons are scattered a large number of times in the comoving frame, and the transfer of energy and momentum between the core and the sheath occurs diffusively{}{. The distance over which photons diffuse to mediate this transition can be estimated by noting that, if the flow is moving predominantly radially, photons are scattered a number of times $N \sim r/\lambda_{\rm mfp} = r\rho'\Gamma \kappa$ during the time over which a fluid element crosses a radial distance $r$, where $\lambda_{\rm mfp} \simeq 1/(\rho\kappa)$ is the photon mean free path and $\rho = \rho'\Gamma$ is the density of scatterers in the lab frame. The root-mean-square distance in the transverse direction that a given photon travels over this time is then $\sim \lambda_{\rm mfp}\sqrt{N}$, and equating this distance to a characteristic angular scale $r\delta \theta$ yields $\delta \theta \simeq 1/\sqrt{\rho'\Gamma\kappa r}$. We therefore expect that, in the optically thick regime, this characteristic angular scale is the one over which photons mediate the transition between a slower moving sheath and fast-moving core. }{}{This angular scaling can be derived more rigorously by balancing the viscous (diffusive) terms and the inviscid terms in the equations of radiation hydrodynamics (Equation \ref{radhydro}), as this equality is by definition upheld in regions of the flow that are mediated by radiation viscosity; this is also the standard approach for determining the scale over which viscosity is important for determining the dynamics of a fluid (e.g., \citealt{kundu08}). }

Radiation-mediated, structured jets therefore naturally possess a smallness parameter $\delta \theta$, which describes the ``boundary layer'' thickness over which the flow assumes this structured form. Within this boundary layer we expect there to be established a transverse velocity, with a magnitude smaller than that of the radial component by a factor $\sim \delta\theta$, as a consequence of the transfer of momentum to and from the fast-moving core. Such considerations lead to the reduction of the full set of equations of radiation hydrodynamics to a simplified, boundary layer form (see \citealt{arav92, coughlin15a, coughlin15b}), within which the viscous energy-momentum tensor is given by

\begin{equation}
R^{\mu\nu}_{\rm vis} = -\frac{8}{9}\frac{p'}{\rho'\kappa}\left(\Pi^{\mu\sigma}\nabla_{\sigma}U^{\nu}+\Pi^{\nu\sigma}\nabla_{\sigma}U^{\mu}\right)-\frac{1}{\rho'\kappa}\left(U^{\mu}\Pi^{\nu\sigma}+U^{\nu}\Pi^{\mu\sigma}\right)\nabla_{\sigma}p'.
\end{equation}
In such a boundary layer approximation, the equations of radiation hydrodynamics are therefore\footnote{Written covariantly and with the Einstein summation convention, Equation \eqref{radhydro} is deceptively compact, and contains a large number of terms when expanded out (see, e.g., \citealt{greenberg75}). In Appendix \ref{sec:appA} we include a derivation of the entropy and Bernoulli equations (Equations \ref{entropy} and \ref{bernoulli} below) that shows that a number of these terms are sub-dominant in the boundary layer approximation, where there is an ordering of the magnitude of the components of the four-velocity and the derivatives.}

\begin{equation}
\rho'U^{\mu}\nabla_{\mu}U^{\nu}+\nabla_{\mu}\left[4p'U^{\mu}U^{\nu}\right]+g^{\mu\nu}\nabla_{\mu}p' = \nabla_{\mu}\left[\frac{8}{9}\frac{p'}{\rho'\kappa}\left(\Pi^{\mu\sigma}\nabla_{\sigma}U^{\nu}+\Pi^{\nu\sigma}\nabla_{\sigma}U^{\mu}\right)+\frac{1}{\rho'\kappa}\left(U^{\mu}\Pi^{\nu\sigma}+U^{\nu}\Pi^{\mu\sigma}\right)\nabla_{\sigma}p'\right]. \label{radhydro}
\end{equation}
Here we used the continuity equation, written covariantly as

\begin{equation}
\nabla_{\mu}\left[\rho'U^{\mu}\right] = 0, \label{cont0}
\end{equation}
which ensures the conservation of (baryonic) particle number. 

We will assume here that the accretion flow that establishes the structured jet is maintained over many sound-crossing times of the radial scale length. In this case, which should be applicable to hyperaccreting systems such as gamma-ray bursts and tidal disruption events, the outflow effectively reaches a time-steady configuration, and the time dependence in Equation \eqref{radhydro} can be omitted. Also, our assumption of an approximately spherical injection of energy motivates that we analyze solutions to Equation \eqref{radhydro} in spherical coordinates, and we will let the flow be azimuthally symmetric (such that there are no derivatives with respect to the azimuthal angle in Equation \eqref{radhydro} when expanded out).

Within the boundary layer, angular variations occur over the angular range $\delta \theta  \simeq 1/\sqrt{\rho'\kappa\Gamma r}$. It is therefore reasonable to change variables to the normalized polar angle $\theta / \delta \theta$, and to look for solutions that -- to leading order in the boundary layer thickness -- vary as $U^{\theta} \propto \delta\theta$, $U^{r} \propto 1$, $\rho' \propto 1$, and $p' \propto 1$. With this ordering, {}{if we focus on the $\nu = \theta$ component of Equation \eqref{radhydro}, the first two terms on the left-hand side are advective derivatives of the $\theta$-component of the velocity; noting that the advective derivative is $\sim U^{r}\partial/\partial r+U^{\theta}\partial/\partial\theta \sim \mathcal{O}(1)$ in the boundary layer thickness, these two terms are of order $\delta\theta$. Similarly, since $\delta \theta \sim 1/\sqrt{\rho'\kappa\Gamma r}$, the right-hand side of this equation is also of order $\delta \theta$. The pressure gradient, however, is $\partial p'/\partial \theta \sim p'/\delta \theta$, and is therefore much larger than the other terms; the only way to satisfy the $\theta$-component of the momentum equation in this expansion is therefore to maintain}

\begin{equation}
\frac{\partial p'}{\partial \theta} = 0, \label{peq}
\end{equation}
and hence the pressure is constant across the width of the boundary layer {}{to leading order in the boundary layer thickness}. This statement can be interpreted to mean that the boundary layer itself is in internal causal contact, such that the transverse sound speed has had time to communicate pressure equilibrium in the angular direction. Equation \eqref{peq} also implies that we can ignore the conduction term -- the second term in parentheses on the right-hand side of Equation \eqref{radhydro} -- in the other components of Equation \eqref{radhydro}. 

We will let the pressure profile be approximated as a power-law with radius, and -- given Equation \eqref{peq} -- write the comoving pressure as

\begin{equation}
p' = p'_{\rm j}\left(\frac{r}{R}\right)^{-n}, \label{ppow}
\end{equation}
where $n$ is a positive number and $R$ is a radial scale length {}{(not to be confused with the symbol used for the energy-momentum tensor of the radiation field)}. Written in this way, we can always redefine the radial scale $R$ and the pressure normalization $p'_{\rm j}$ so that the length $R$ does not directly affect the solution, which suggests that we can find self-similar solutions to Equation \eqref{radhydro} that depend only on conditions within the boundary layer. {}{In spite of the fact that we can set $R$ and $p'_{\rm j}$ to unity without loss of generality, we maintain the above expression for the pressure (and always specifically include reference to the scale length $R$ when referring to the spherical radius $r$) to avoid potential confusion with units. }

Restricting the pressure to vary as a power-law in radius clearly limits the possible solutions that we can find to the equations of radiation hydrodynamics. However, as we discuss at length in Section \ref{sec:implications}, the physical conditions that arise in both jetted tidal disruption events and gamma-ray bursts are likely conducive to the formation of pressure profiles that at least roughly conform to power laws. At the very least, adopting a simple power-law prescription for the pressure profile that drives the outflow will permit self-similar and analytically tractable solutions, which will engender fundamental understanding as to the bulk features of such relativistic and structured flows. 

While we could work directly with the individual components of Equation \eqref{radhydro}, it is more straightforward to deal with the entropy equation and the Bernoulli equation, which are obtained by taking the contraction of Equation \eqref{radhydro} with $U_{\nu}$ and $\Pi^{r}_{\,\,\nu}$ respectively. Using the $\theta$-independence of the pressure from Equation \eqref{peq} {}{and making a modest number of algebraic manipulations (in Appendix \ref{sec:appA} we give an abridged derivation for the interested reader)}, these equations are

\begin{equation}
U^{\mu}\partial_{\mu}\ln\left(\frac{p'}{\left(\rho'\right)^{4/3}}\right) = \frac{8}{27}\frac{1}{\rho'\kappa}\frac{1}{r^2}\left(\frac{1}{\Gamma}\frac{\partial}{\partial \theta}\left[\Gamma v_{\rm r}\right]\right)^2, \label{entropy}
\end{equation}
\begin{equation}
\rho'U^{\mu}\partial_{\mu}\left[\Gamma\left(1+4\frac{p'}{\rho'}\right)\right]= \frac{8}{9}\left\{\frac{p'}{\rho'\kappa}\frac{1}{\Gamma}\left(\frac{1}{r\Gamma}\frac{\partial}{\partial \theta}\left[\Gamma v_{\rm r}\right]\right)^2+v_{\rm r}\frac{1}{r^2\theta}\frac{\partial}{\partial \theta}\left[\frac{p'}{\rho'\kappa}\theta\frac{\partial}{\partial \theta}\left[\Gamma v_{\rm r}\right]\right]\right\}. \label{bernoulli}
\end{equation}
Here the radial component of the three velocity is given by $v_{\rm r}$, and to order $\delta\theta^2$ the Lorentz factor is $\Gamma  = (1-v_{\rm r}^2)^{-1/2}$. As one would expect, these equations demonstrate that the specific entropy $p'/(\rho')^{4/3}$ and the Bernoulli parameter $\Gamma\left(1+4p'/\rho'\right)$ are conserved along flow lines in inviscid regions of the flow. The fact that the right-hand side of Equation \eqref{entropy} is positive-definite also implies that viscous dissipation serves only to increase the entropy, in agreement with the second law of thermodynamics. {}{We also used the small-angle approximation here and set $\sin\theta \simeq \theta$, as we expect the outflow to be confined to the region $\delta\theta \ll 1$. Finally, the continuity equation (in the same small-angle approximation) reads}

\begin{equation}
\frac{1}{r^2}\frac{\partial}{\partial r}\left[r^2\rho'\kappa\Gamma v_{\rm r}\right]+\frac{1}{r\theta}\frac{\partial}{\partial \theta}\left[\theta\rho'\kappa\Gamma v_{\theta}\right] = 0. \label{cont0}
\end{equation}
For convenience (in the next section) we multiplied this equation by the opacity $\kappa$, assumed to be constant in the comoving frame. Equations \eqref{entropy} -- \eqref{cont0}, in addition to the conservation of the norm of the four velocity, constitute the three equations for the three unknowns $\Gamma v_{\rm r}$, $v_{\theta}$, and $\rho'$. In the next section we derive self-similar solutions to these equations that describe the propagation of a relativistic jet powered by a hyperaccreting source.

\section{Self-similar, Relativistic, Radiation-dominated Jets}
\label{sec:jets}
\subsection{Definitions and Equations}
\label{sec:equations}
We solve the continuity equation \eqref{cont0} by introducing the stream function $\psi$:

\begin{equation}
\rho'\kappa r^2\theta\Gamma v_{\rm r} = \frac{\partial \psi}{\partial \theta}, \,\,\, \rho'\kappa r\theta \Gamma v_{\theta} = -\frac{\partial \psi}{\partial r}. \label{psieq}
\end{equation}
We assume that along its axis the jet four-velocity varies as a power-law in radius, such that $\Gamma v_{\rm r} \propto r^{m}$ at $\theta = 0$ with $m$ a pure number. We expect $m$ to be constrained by the fact that, if the energy driving the outflow is sourced at the hyperaccreting black hole, the flux of energy through the solid angle of the outflow at any given radius is a conserved quantity. We also expect the jet Lorentz factor to be regulated by the fact that a pressure gradient is trying to accelerate the outflow, but as the fluid expands outward into its surroundings, the total amount of inertia -- primarily in the form of radiation for the solutions considered here -- contained within the outflow increases; because the energy flux is conserved, this increase in the inertia must be balanced by a reduction in the Lorentz factor. We will show below that the value of $m$ can be determined from energy conservation and the assumption of self-similarity once we specify the radial dependence of the pressure and density by enforcing energy conservation, but for now we leave it as an unspecified parameter.

As we argued above, there is a characteristic angular scale $\delta\theta \simeq 1/\sqrt{\rho'\kappa \Gamma r}$ over which we expect the jet to vary due to the presence of radiation viscosity. A change of variables to the self-similar variable $\xi = \theta/\delta\theta$ therefore naturally establishes this small parameter as the one that primarily determines the angular width of the outflow. Given these considerations, we look for solutions that vary self-similarly in this variable, and in particular we parameterize the stream function as

\begin{equation}
\psi = rf(\xi),
\end{equation}
\begin{equation}
\xi \equiv \Gamma_{\rm j}v_{\rm j}\left(\frac{r}{R}\right)^{m}\int_0^{\theta}\rho'\kappa r \theta d\theta.
\end{equation}
Here $\Gamma_{\rm j}v_{\rm j}$ is the radial component of the jet four velocity along the axis at the scale radius $R$. With this parameterization, Equation \eqref{psieq} shows that the radial component of the four-velocity is

\begin{equation}
\Gamma v_{\rm r} = \Gamma_{\rm j}v_{\rm j}\left(\frac{r}{R}\right)^{m}\frac{df}{d\xi}. \label{gammadef}
\end{equation}
Note that, even though we are defining the self-similar variable implicitly in terms of the density, it varies as $\xi \sim \left(\theta/\delta\theta\right)^2$. Also from Equation \eqref{psieq}, the angular component of the jet four velocity satisfies

\begin{equation}
r\theta\rho'\kappa \Gamma v_{\theta} = -\left(f+r\frac{\partial\xi}{\partial r}\frac{df}{d\xi}\right). \label{vtheta}
\end{equation}
To be consistent with the definition that the jet velocity be $v_{\rm j}$ along the axis, we have the boundary condition $df/d\xi(\xi = 0) = 1$. By symmetry we also impose that the angular component of the velocity go to zero along the axis, and hence we have the additional boundary condition $f(0) = 0$. We further assume that the density varies self-similarly as

\begin{equation}
\rho' = \rho'_{\rm j}\left(\frac{r}{R}\right)^{-q}g(\xi),
\end{equation}
where $\rho'_{\rm j}$ is a characteristic density within the jetted outflow and $q$ is a pure number. As we will see below, the density of scatterers along the axis may go to zero depending on the run of density and pressure with radius, and hence we do not necessarily have the freedom to set $g(0) = 1$. 

We can now insert the above expressions for the velocity and the density into the entropy and Bernoulli equations, \eqref{entropy} and \eqref{bernoulli}. We will assume that the internal energy of the gas is relativistic, such that the enthalpy of the radiation dominates the rest-mass energy of the scatterers, which is an appropriate limit in the context of hyperaccreting systems\footnote{It is not always necessary to take this limit to recover self-similar solutions, as we do in the next section; see \citealt{arav92} and \citet{coughlin15a, coughlin15b}, where the additional parameter of the ratio of the rest-mass energy to the radiation energy density (in the comoving frame) is maintained.}. In this limit the explicit dependence on the magnitude of the radiation energy density $e'$ drops out of the equations of radiation hydrodynamics, as each term (including the viscous ones) depends linearly on this quantity. Adopting this limit also shows that the Bernoulli parameter ( $=\Gamma\left(1+4p'/\rho'\right)$, the quantity in brackets on the left-hand side of Equation \ref{bernoulli}) becomes

\begin{equation}
\Gamma\left(1+\frac{4p'}{\rho'}\right) \simeq \frac{4p'\Gamma}{\rho'} = \frac{4p'_{\rm j}}{\rho'_{\rm j}}\left(\frac{r}{R}\right)^{q-n}\frac{1}{g}\sqrt{1+\Gamma_{\rm j}^2v_{\rm j}^2\left(\frac{r}{R}\right)^{2m}\left(\frac{df}{d\xi}\right)^2},
\end{equation}
while the specific entropy is

\begin{equation}
s = \ln\left(\frac{p'}{(\rho')^{4/3}}\right) = \left(\frac{4}{3}q-n\right)\ln r-\frac{4}{3}\ln g.
\end{equation}
Using the above expressions for the four-velocity in terms of the stream function, it also follows that the advective derivative can be written generally as

\begin{equation}
U^{\mu}\partial_{\mu} = \frac{1}{r}\Gamma_{\rm j}v_{\rm j}\left(\frac{r}{R}\right)^{m}\left(r\frac{df}{d\xi}\frac{\partial}{\partial r}-f\frac{\partial}{\partial \xi}\right).
\end{equation}
In the next subsection we present analytic (and closed-form) solutions to the fluid equations that make the ultra-relativistic approximation, such that $v_{\rm j} \simeq 1$ and $\Gamma v \simeq \Gamma \gg 1$. 

\subsection{Ultra-relativistic, energy-conserving, self-similar solutions}
\label{sec:ultra}
When the flow is ultra-relativistic, we have 

\begin{equation}
\Gamma v \simeq \Gamma = \Gamma_{\rm j}\left(\frac{r}{R}\right)^{m}\frac{df}{d\xi},
\end{equation}
and the Bernoulli parameter simplifies to

\begin{equation}
\frac{4p'\Gamma}{\rho'} = \frac{4p'_{\rm j}\Gamma_{\rm j}}{\rho'_{\rm j}}\left(\frac{r}{R}\right)^{m-n+q}\frac{1}{g}\frac{df}{d\xi}.
\end{equation}
It is also apparent that the first term on the right-hand side of Equation \eqref{bernoulli} is smaller than the second term by a factor of $\sim 1/\Gamma_{\rm j}^2$, and may therefore be ignored in this approximation. Inserting our definitions of the fluid variables into the fluid equations, we can show that they become

\begin{equation}
\left(m-n+q\right)\frac{dh}{d\xi}\left(\frac{df}{d\xi}\right)^2-f\frac{d}{d\xi}\left[\frac{dh}{d\xi}\frac{df}{d\xi}\right] = \frac{4}{9}\frac{d}{d\xi}\left[h\frac{d^2f}{d\xi^2}\right], \label{sstot1}
\end{equation}
\begin{equation}
\left(q-\frac{3n}{4}\right)\frac{dh}{d\xi}\frac{df}{d\xi}-f\frac{d^2h}{d\xi^2} = \frac{4}{9}h\left(\frac{df}{d\xi}\right)^{-2}\left(\frac{d^2f}{d\xi^2}\right)^2. \label{sstot2}
\end{equation}
Here we defined the function

\begin{equation}
h(\xi) \equiv \int_0^{\xi}\frac{d\tilde{\xi}}{g(\tilde{\xi})},
\end{equation}
and we inverted the definition of $\xi(\theta)$ to yield

\begin{equation}
\frac{1}{2}\theta^2 = \frac{1}{\Gamma_{\rm j}\rho'_{\rm j}\kappa R}\left(\frac{r}{R}\right)^{q-m-1}h(\xi). \label{thetaofxi}
\end{equation}
As we noted above, we expect the energy powering the jet to be provided by the hyperaccreting object at the origin, and hence the flux of energy through the jetted outflow at any given radius should be a conserved quantity. In the ultra-relativistic and small-angle limits, the energy flux through the solid angle of the jet is given by

\begin{equation}
\dot{E} = \int T^{0 r} r^2\sin\theta d\theta d\phi \simeq \int 8\pi p'\Gamma^2 r^2 d\left(\frac{1}{2}\theta^2\right) \propto r^{m-n+q+1}, \label{Edot1}
\end{equation}
where we used the self-similar scalings for the radiation pressure and the Lorentz factor, and we changed variables from $\theta$ to $\xi$ by using Equation \eqref{thetaofxi} (see also Equation \ref{Edot} below where the numerical and dimensional factors are retained). If this expression is to be independent of radius we therefore require

\begin{equation}
m-n+q = -1.
\end{equation}
Note that it is precisely this combination of parameters that multiplies the first term in Equation \eqref{sstot1}; with this number set to $-1$, it can be verified that the solution to Equations \eqref{sstot1} and \eqref{sstot2} is

\begin{equation}
f = \frac{4}{9}\frac{1}{\beta}\ln\left(1+\frac{9}{4}\beta\xi\right), \,\,\, h = \frac{1}{\beta}f^{\beta}, \,\,\, \beta \equiv q-\frac{3n}{4}+1. \label{exsols}
\end{equation}
This solution satisfies the boundary conditions on the function $f$, namely $f(0) = 0$ and $df/d\xi(0) = 1$. Since $h$ is an integral over the density starting from the axis, we require that $h(0) = 0$, which results in the constraint that $\beta > 0$. The self-similar density is given by

\begin{equation}
g = \left(\frac{dh}{d\xi}\right)^{-1} = f^{1-\beta}\left(\frac{df}{d\xi}\right)^{-1}.
\end{equation}
Because the entropy increases in the presence of shear, we expect the density to decline as we approach the axis of the jet; to be consistent with this expectation we require that $\beta \le 1$. 

\subsection{Jet Properties}
\label{sec:properties}
Here we delimit and explore the main features of the self-similar, ultra-relativistic jet solutions derived above and their dependence on the pressure power-law index $n$ and the density power-law index $q$. 

\subsubsection{Angular dependence and structure}
\label{sec:angular}
The solutions given by Equation \eqref{exsols} are written in terms of the self-similar variable $\xi$, which itself is defined in terms of the density. We can recover the physical variable $\theta$ by returning to Equation \eqref{thetaofxi}, using the solution for $h$, and inverting the expression, which yields

\begin{equation}
\xi = \frac{4}{9\beta}\left\{\exp\left[{\frac{9}{4}\beta\left(\frac{\beta}{2}\left(\frac{\theta}{\delta\theta(r)}\right)^{2}\right)^{1/\beta}}\right]-1\right\}, \label{xigauss}
\end{equation}
where $\delta\theta(r)$ conforms to the scaling derived above for the boundary layer thickness:

\begin{equation}
\delta\theta(r) \equiv \left(\rho'\kappa \Gamma r\right)^{-1/2} = \left(\rho'_{\rm j}\kappa\Gamma_{\rm j}R\right)^{-1/2}\left(\frac{r}{R}\right)^{q-n/2}. \label{dtheta}
\end{equation}
In the last line we used the constraint from energy conservation that $m = n-q-1$. 

\begin{figure} 
   \centering
   \includegraphics[width=0.485\textwidth]{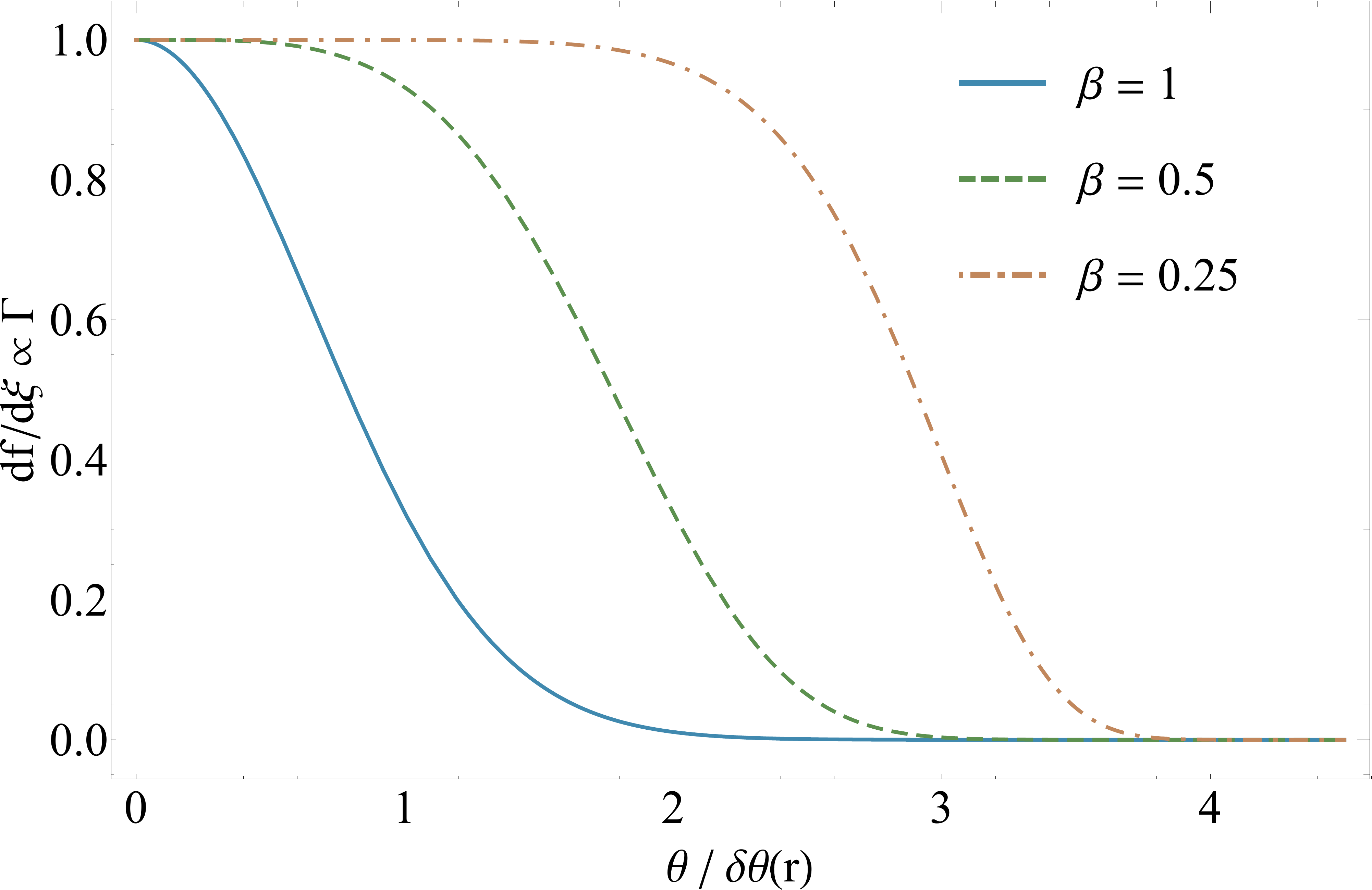} 
   \includegraphics[width=0.5\textwidth]{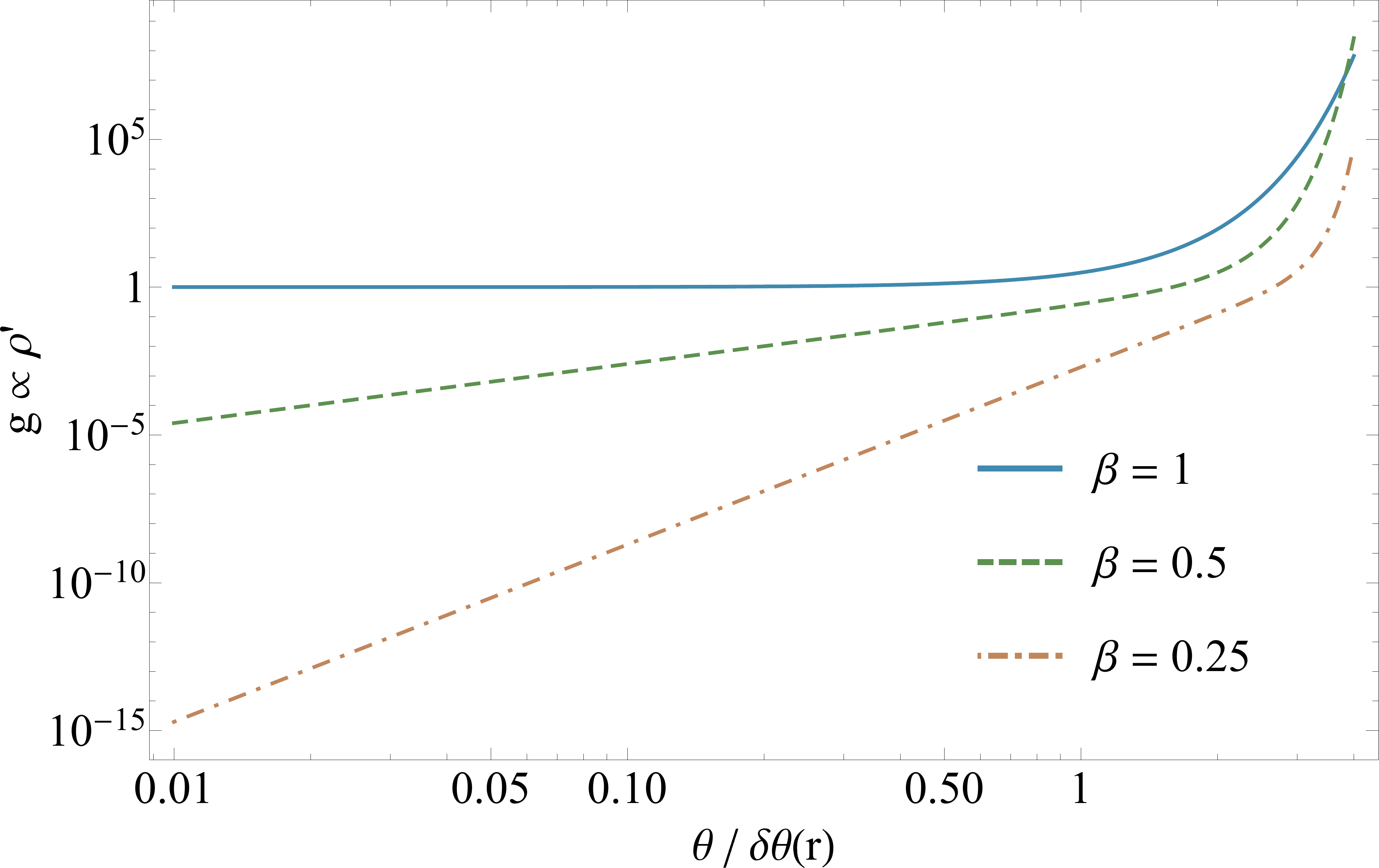} 
   \caption{Left: The variation of the Lorentz factor within the outflow, normalized by the jet Lorentz factor along the axis, as a function of angle from the jet axis. When $\beta = 1$ the solution is a Gaussian, and as $\beta$ decreases toward zero the variation becomes increasingly top-hat-like (see Equation \ref{xigauss} for the relationship between the self-similar variable $\xi$ and the polar angle $\theta$); the velocity therefore rapidly becomes non-relativistic outside the scale angle $\delta\theta(r)$, which corresponds to the characteristic mean-free path across the jet (see Equation \ref{dtheta} for the definition of this angle and its radial dependence). Right: The normalized density of scatterers within the jet as a function of polar angle (note that this panel is on a log-log scale). When $\beta = 1$ the solution levels off to a constant value along the axis, and as $\beta$ decreases the flow becomes increasingly evacuated in the inner regions. The density increases dramatically outside of $\theta \simeq \delta\theta$, and thus most of the mass is concentrated at the angular extremities of the jet.  }
   \label{fig:gammarho}
\end{figure}

Figure \ref{fig:gammarho} illustrates the Lorentz factor of the outflow (left panel) and the comoving density (right panel) as functions of polar angle normalized by the boundary layer thickness $\delta\theta$; note that the right panel is on a log-log scale. The different curves are for different values of $\beta$, with larger $\beta$ corresponding to less radial variation in the specific entropy (i.e., as $\beta$ decreases to zero the specific entropy is more radially concentrated near the origin, presumed to coincide with a hyperaccreting object that is responsible for the energy injection that drives the outflow). When the flow is isentropic, such that $q=3n/4$ and $\beta = 1$, Equation \eqref{xigauss} combined with the fact that the Lorentz factor of the flow satisfies $\Gamma \propto df/d\xi = (1+9\beta\xi/4)^{-1}$ shows that the radial component of the four-velocity of the jet declines as a Gaussian from the axis. The Lorentz factor is thus approximately constant at $\sim \Gamma_{\rm j}$ for $\theta \lesssim \delta\theta$ and very rapidly approaches zero for $\theta \gtrsim \delta\theta$, which is apparent from the left panel of Figure \ref{fig:gammarho}. When the specific entropy is maximized toward the black hole, such that $\beta < 1$, the Lorentz factor falls off even more steeply outside of $\sim \delta\theta$. 

The right panel of this figure illustrates the solution for the density for the same set of $\beta$. In the limit of small-$\xi$, the relationship between the self-similar density and the function $f$ shows that $g \simeq (\theta/\delta\theta)^{2(1-\beta)/\beta}$. Therefore, when $\beta = 1$ and the flow is isentropic with radius, the density approaches a constant along the axis; otherwise (for $\beta < 1$) the density goes to zero as a power-law in $\theta$ as we near the jet axis, with the power-law index increasing with decreasing $\beta$. At large $\theta/\delta\theta(r)$, the density approximately \emph{increases} as a Gaussian, which follows from the analytic relationship between the functions $f$ and $g$ and can be seen qualitatively from the right-hand panel of Figure \ref{fig:gammarho}. These jetted solutions are therefore nearly evacuated in the interior, and very rapidly become dominated by rest mass outside the self-similar angle $\delta\theta$. 

{}{We note that the second derivative of $f$ with respect to $\xi$ reaches its maximum value along the axis, given by $d^2f/d\xi^2(\xi = 0) = -9\beta/4$. However, the shear within the outflow (in the angular direction) is proportional to $\partial\Gamma/\partial \theta$, which is zero along the axis -- as it must be by symmetry -- and at large angles from the axis. Therefore, the shear within the outflow is maximized at the angle $\sim \delta\theta$ where the outflow merges with the ambient gas.}

\subsubsection{Collimation and Causal Connectedness}
\label{sec:collimation}
From Equation \eqref{xigauss} it follows that $\xi$ is a function of $\theta/\delta\theta(r)$, where -- with the relationship $m = n-q-1$ enforced by energy conservation -- the boundary layer thickness $\delta\theta(r)$ is given by

\begin{equation}
\delta\theta(r) \propto r^{q-n/2}. \label{collcond}
\end{equation}
This scaling illustrates that the degree of jet collimation depends on the relative scaling between the pressure and density, and that the jet is successfully collimated by the ambient medium only if $q \le n/2$. Since the Lorentz factor of the jet is proportional to the function $f$, which declines monotonically from the jet axis ($\xi = 0$), surfaces on which the Lorentz factor varies as a fixed ratio of the Lorentz factor along the axis coincide with surfaces of constant $\xi$. Thus, when $q < n/2$, maintaining a constant $\xi$ corresponds to moving to smaller angles from the axis as we move to larger radii, and hence the jet is \emph{collimated} by the ambient medium: surfaces on which the Lorentz factor is larger (as a function of the axial value) are increasingly confined to the jet axis. The behavior is inverted when $q > n/2$, with relativistic flow persisting to larger angles as we move farther down the jet axis, and the jet is de-collimated in this regime. Finally, if $q \equiv n/2$, the jet is marginally collimated and the flow lines are conical. 

\begin{figure}
   \centering
   \includegraphics[width=0.3\textwidth]{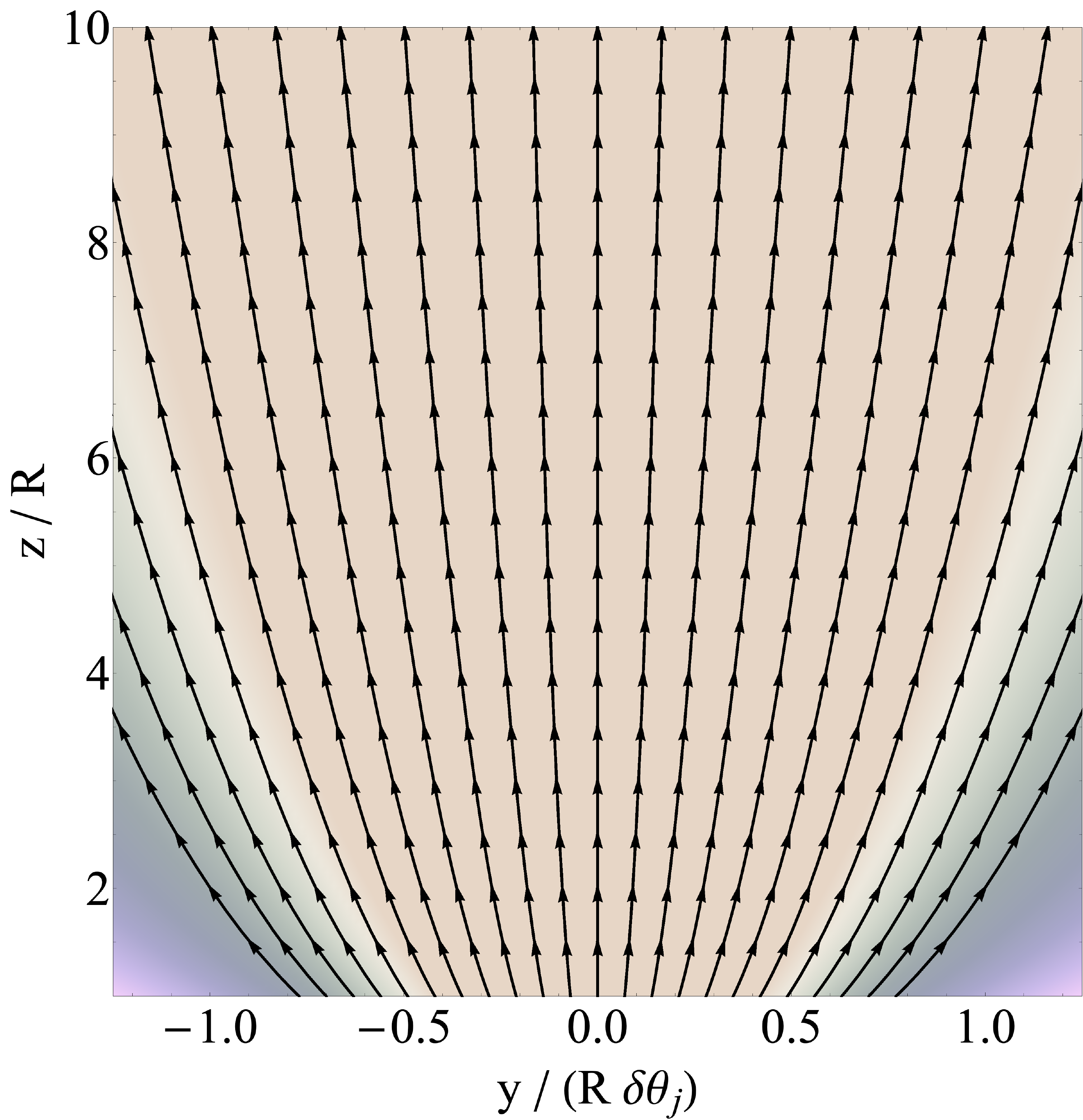} 
  \includegraphics[width=0.3\textwidth]{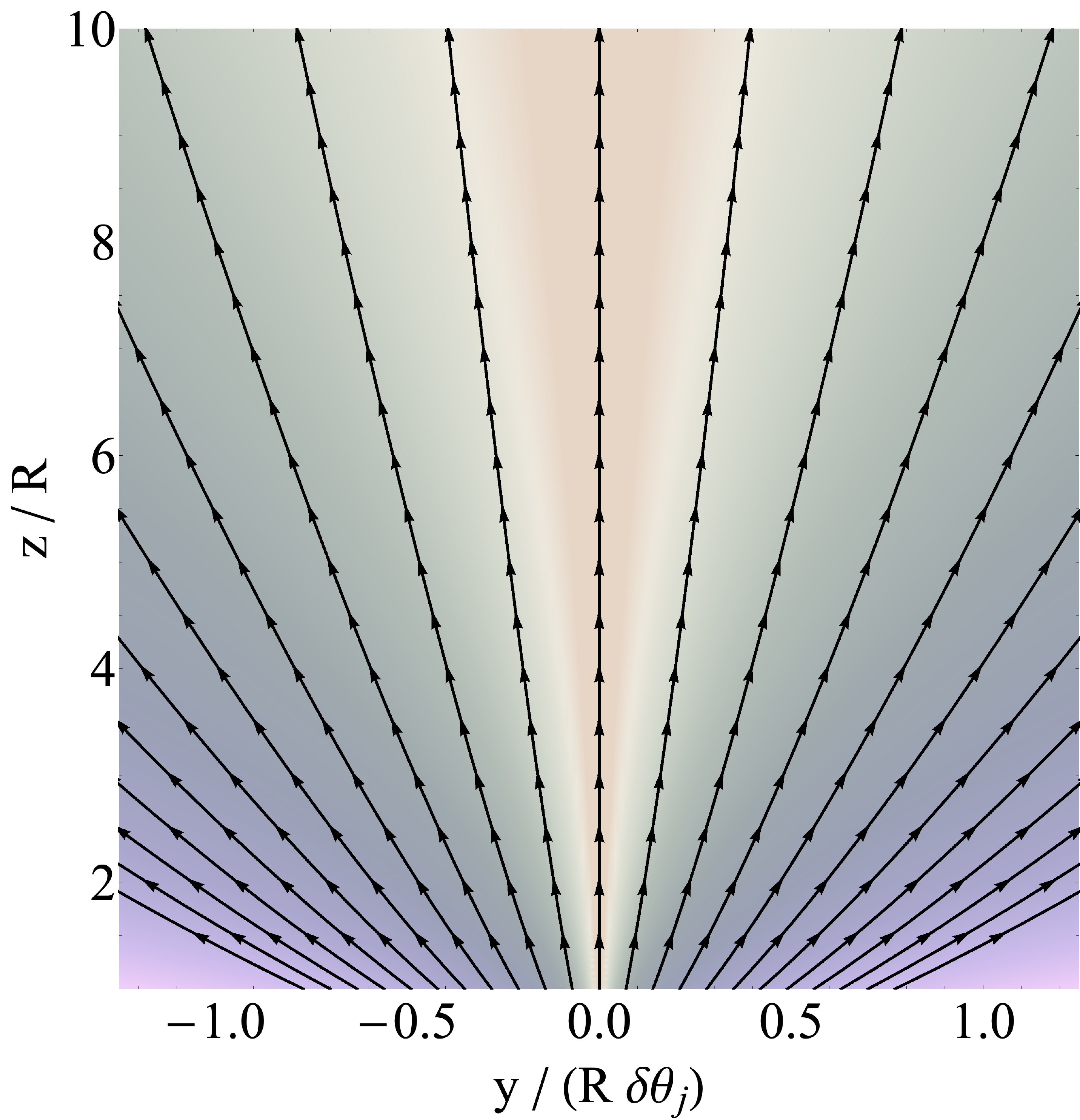} 
  \includegraphics[width=0.39\textwidth]{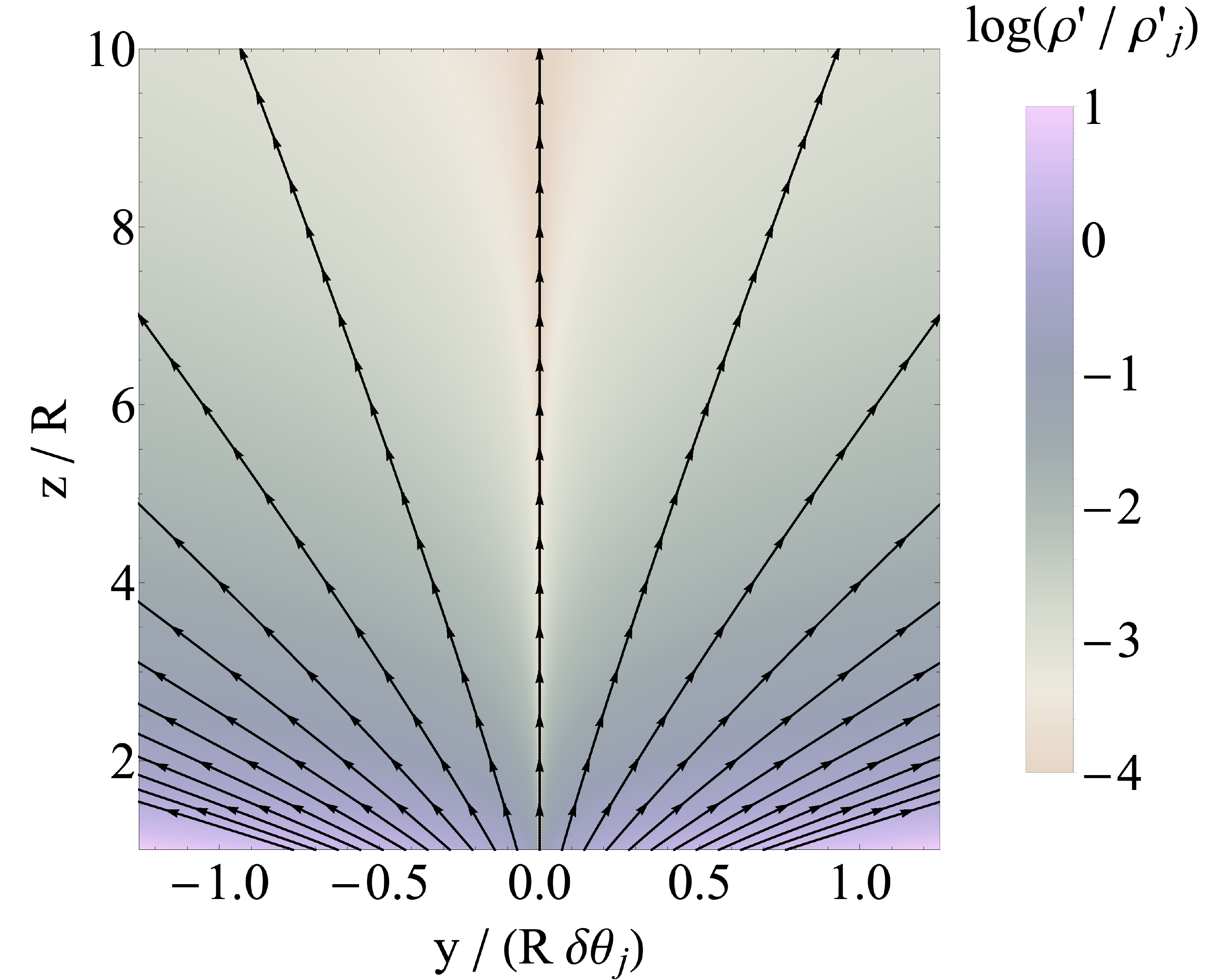} 
   \caption{Streamlines of the four-velocity of the ultra-relativistic jet solutions (black lines; arrows indicate the directionality of the flow). For definitiveness we show solutions where the radial power-law index of the density, $q$, is related to that of the pressure, $n$, via $q = n-1$, so that the Lorentz factor of the jet is constant along the jet axis (i.e., the radial power-law index of the Lorentz factor, $m$, is zero). The colors indicate the (base 10) logarithm of the density, as shown in the color bar. The length along the jet axis is normalized by the scale radius $R$, and the length perpendicular is relative to the boundary layer thickness $R\delta\theta_{\rm j}$, where $\delta \theta_{\rm j} = (\rho'_{\rm j}\Gamma_{\rm j}\kappa R)^{-1/2}$ is the angular width of the boundary layer at the scale radius $R$. The left panel has $n = 1$, and the jet remains collimated with the streamlines closing up toward the axis; the middle panel has $n = 2$, in which case the streamlines asymptotically lie on cones with a constant opening angle; and the right panel has $n = 3$, the jet is not collimated by the ambient medium, and the streamlines open up to larger angles farther down the axis of the jet.}
   \label{fig:streams}
\end{figure}

Solid black lines in Figure \ref{fig:streams} illustrate the streamlines of the four-velocity, calculated from the vector equation for the characteristics along the flow, $dx^{\mu}/d\tau = U^{\mu}$, where $x^{\mu}$ are the coordinates ($r$ and $\theta$) and $U^{\mu}$ is the four-velocity (the $\theta$-component of the four-velocity is calculated from Equation \ref{vtheta}; see also \ref{vtheta2} below for an algebraically simplified expression); arrows indicate the directionality of the flow. For the sake of concreteness, we picked solutions that have a constant Lorentz factor along the jet axis, such that $m = 0$ and hence the radial power-law indices of the pressure and density, $n$ and $q$, are related via $n = q+1$. The left, middle, and right panels have $n = 1$, 2, and 3, respectively, meaning that the jet in the left (right) panel is collimated (uncollimated) and the streamlines converge toward (diverge from) the axis as we move to larger distances from the origin. The middle panel is marginally collimated, and the streamlines maintain a constant jet opening angle far from the origin.

The colors in this figure indicate the base-10 logarithm of the density, as shown by the color bar on the right. For these combinations of parameters, the left panel has $\beta = 1/4$, the middle panel has $\beta = 1/2$, and the right panel has $\beta = 3/4$. As $\beta$ decreases, the density declines as a steeper power-law in angle toward the jet axis, as shown in the right panel of Figure \ref{fig:gammarho} above. Thus, the inner, fast-moving core of the solution with $n = 1$ (left panel) possesses a much larger region that is nearly evacuated compared to that with $n = 3$. Also note that the density is independent of radius in the left-hand panel, while it declines as a function of radius for the middle ($\rho' \propto r^{-1}$) and right ($\rho' \propto r^{-2}$) panels, and this decline in the density is also reflected in the coloration of the panels.

Interestingly, Equation \eqref{collcond} illustrates that an isentropic jet, which has $q = 3n/4$ (or $\beta = 1$ in terms of the parameters introduced in Section \ref{sec:ultra}), is \emph{not collimated}, and instead the flow lines diverge away from the axis as $\propto r^{n/4}$. The successful collimation of the jet therefore \emph{requires} a more concentrated radial profile of specific entropy. In particular, since the condition for collimation is that $q \le n/2$, we have that $\beta$ must satisfy $\beta \le 1-n/4$ for the jet to be collimated by the ambient medium. Thus, the successful collimation of the outflow requires a steeper specific entropy profile when the pressure profile of the ambient gas is steeper (i.e., $\beta$ must decrease below 1 as $n$ increases above 0). 

Causal connectedness across the jet is maintained provided that

\begin{equation}
\Gamma\delta\theta \lesssim 1. \label{causcond}
\end{equation}
When this inequality is satisfied, the transverse sound speed is larger than the rate at which the jet expands sideways into its surroundings, and we expect our scalings and the constancy of the comoving pressure with angle to be maintained (also note that this inequality was assumed to hold in the entropy equation; see the discussion in Appendix \ref{sec:appA}). Using the value of $m$ in terms of the ambient pressure and density power-law indices and the boundary layer thickness $\delta\theta$ then shows that causal connectedness will be maintained when $n \le 2$. This condition on causal connectedness is completely independent of the variation in the density profile, and implies that these jets will continue to be causally connected provided that the pressure of the ambient medium falls off less steeply than $\propto r^{-2}$, i.e., when the pressure falls off less (more) steeply than $r^{-2}$, the product $\Gamma \delta\theta$ decreases (increases) as a function of radius. When causal connectedness is not maintained, it is possible that the supersonic expansion of the jet results in the formation of shocks that then serve as a distinct mechanism to recollimate the flow, introducing additional length scales into the problem that violate the self-similar approach adopted here (we refer the reader to, e.g., \citealt{bromberg07, kohler12, kohler15} for more detailed discussions of shock collimation).

\subsubsection{Acceleration and Stability}
As we found above, the conservation of energy flux throughout the jetted outflow requires that the acceleration parameter of the jet, $m$, vary according to $m = n-q-1$. For positive $m$, the jet accelerates along the axis (and along any surface of constant $\xi$), while the outflow decelerates for negative $m$, and hence the condition necessary for jet acceleration is that $n > q+1$. Physically, this statement can be interpreted to mean that when the pressure of the ambient medium falls off sufficiently rapidly, the increase in the inertia contained in the outflow as we move outward in radius does not compensate for the force provided by the pressure gradient. On the other hand, when the pressure gradient is weak (or, indeed, absent), the increase in the inertia dominates the accelerating effect of the pressure profile of the gas, and the outflow decelerates\footnote{Note that when there is no pressure gradient ($n = 0$), the only solutions are decelerating ($m < 0$), and that steeper density gradients ($q > 0$) result in a \emph{more rapidly decelerating} jet. This latter relationship arises from the fact that the characteristic opening angle of the jet $\delta\theta$ widens as the density profile steepens, which results in a more dramatic increase in the total inertia contained in the outflow}.

When there is no pressure gradient and the ambient medium is characterized by a constant density, so that $q = n = 0$, the outflow decelerates according to $\Gamma \propto r^{-1}$ and represents a ``freely expanding'' jet within a homogeneous medium. An analogous situation was considered in the relativistic, compressible limit by \citet{coughlin15b}, who adopted a Cartesian geometry\footnote{In this case the flow is infinite and unchanging in the plane-parallel $x$-direction, the flow decelerates along the $z$-axis, and the variation in the Lorentz factor occurs over the Cartesian boundary layer thickness $\delta y/z \sim (\rho'\kappa \Gamma z)^{-1/2}$} and found that the flow decelerates according to $\Gamma v \propto z^{-1/4}$, in agreement with the result for incompressible, non-relativistic flow (e.g., \citealt{kundu08}). This rate of deceleration is clearly much less rapid than the one derived here in spherical coordinates, and this difference arises from the fact that in spherical coordinates the increase in the cross-sectional area of the jet grows as the \emph{square} of the boundary layer thickness. In contrast, the plane-parallel approximation implies that the area of the outflow grows only as a single power of the boundary layer thickness, as the flow is infinite (and unchanging) in the $x$-direction. Thus, and as argued heuristically in Section 5 of \citet{coughlin15b}, the fact that the area of the jet increases more rapidly along the axis in spherical coordinates, and the corresponding larger rate of increase of the inertia contained in the flow, results in the more rapid deceleration of the jet. 

Free-streaming, decelerating jets in the plane-parallel limit have been found to exhibit some degree of dynamical instability, and the laminar nature of the flow tends to break down and is replaced by a more turbulent transition between the fast outflow and the stationary gas. As noted in \citet{kundu08}, this instability is related to the fact that the self-similar solution for the velocity possesses an inflection point, which is a necessary (but not sufficient) condition for dynamical instability (Rayleigh's criterion). In particular, the non-relativistic, incompressible solution has the analytic solution $v_{\rm z} \propto \textrm{sech}^2(\xi)$ for the $z$-component of the velocity, and varies from concave ($\partial^2v_{\rm z}/\partial\xi^2 < 0$) at small $\xi$ to convex ($\partial^2v_{\rm z}/\partial\xi^2 > 0$) at large $\xi$. Interestingly, while the analytic solution derived here for the radial component of the velocity has an inflection point in terms of the physical polar angle (e.g., the left panel of Figure \ref{fig:gammarho}; see also the discussion at the end of Section \ref{sec:angular}), the solution is actually \emph{convex} in terms of the self-similar variable and possesses no such inflection point. The stability of the flow also depends not on the physical variable but on the self-similar one, as the self-similar transformation renders the linear, perturbed equations separable in both a space-like (i.e., $\xi$) and time-like variable and hence conducive to an eigenmode decomposition (e.g., \citealt{coughlin19}). While a complete analysis of the stability of the flows found here is outside the scope of the present, exploratory investigation, these considerations suggest that the solutions found here may be more stable than their non-relativistic, planar, incompressible counterpart. 

It is interesting to note that when $n = q+1$, the pressure gradient precisely balances the increase of inertia along the jet axis to yield a constant Lorentz factor ($m = 0$), and self-similar solutions can be recovered for \emph{arbitrary} Lorentz factors and the ultra-relativistic approximation does not need to be made. We analyze this case in more detail in Appendix \ref{sec:trans} and show that the solutions terminate at a finite value of the self-similar variable in a way that is consistent with more heuristic arguments in the next subsection.

\subsubsection{Energy Flux and the Breakdown of Self-similarity}
\label{sec:energy}
The radial scaling of the energy flux derived in Equation \eqref{Edot1} was used to constrain the radial power-law index of the jet Lorentz factor along the axis. Maintaining all of the coefficients and using the analytic solutions for the self-similar functions, we can show straightforwardly that Equation \eqref{Edot1} becomes

\begin{equation}
\dot{E} = \frac{8\pi p'_{\rm j}\Gamma_{\rm j}Rc}{\rho'_{\rm j}\kappa}\int_0^{\infty}\left(\frac{df}{d\xi}\right)^{3}f^{\beta-1}d\xi \simeq \frac{8\pi p'_{\rm j}\Gamma_{\rm j}Rc}{\rho'_{\rm j}\kappa}\frac{1}{\beta}\left(\frac{2}{9}\right)^{\beta}, \label{Edot}
\end{equation}
where we reintroduced the factor of $c$ for clarity. The integral in this expression cannot be evaluated analytically for arbitrary $\beta$, but it is apparent that the integrand is largest for small-$\xi$, and thus the total energy of the outflow is dominated by the fast-moving, relativistic core. The value of the integral also increases with decreasing $\beta$, and diverges as $\propto 1/\beta$ in the limit that $\beta \rightarrow 0$. In the last line we replaced the integral by the approximation $(2/9)^{\beta}\times1/\beta$, which is exact when $\beta = 1$ and in the limit that $\beta \rightarrow 0$, and provides a good fit in between (the maximum difference between this approximation and the exact integral is at the $\sim 10\%$ level when $\beta \simeq 0.5$, and this approximation systematically overpredicts the true value). Also note that Equation \eqref{Edot} only depends on the values of the density and pressure power-law indices through $\beta$ ($ = q-3n/4+1$), and provided that $\beta$ is not too close to zero, the dependence is not strong.

We can relate the energy flux in Equation \eqref{Edot} to the energy injection rate that arises from the accretion luminosity $L_{\rm acc}$ as material accretes onto the compact object, so that

\begin{equation}
\dot{E} \simeq L_{\rm acc} \equiv \frac{4\pi GMc}{\kappa}\times \ell. 
\end{equation}
Here we parameterized the accretion luminosity in terms of the Eddington luminosity $4\pi GM c/\kappa$, i.e., the ratio of the accretion luminosity to the Eddington luminosity is given by the factor $\ell$. Using the expression for $\dot{E}$ in terms of the jet parameters \eqref{Edot} in this relation, letting the scale radius $R$ be on the order of the gravitational radius $GM/c^2$, and rearranging then shows

\begin{equation}
\frac{4p'_{\rm j}\Gamma_{\rm j}}{\rho'_{\rm j}c^2} \simeq 2\beta\left(\frac{2}{9}\right)^{-\beta}\ell. \label{bernofN}
\end{equation}
The left-hand side of this expression is the relativistic Bernoulli parameter at the base of the outflow, and if our analysis is to be self-consistent (i.e., such that we are justified in dropping the rest-mass energy of the outflow in comparison to this term), it should be much greater than one. This result shows that the Bernoulli parameter is indeed much greater than one when the Eddington ratio of the compact object is large ($\ell \gg 1$), which is a consistency check on these solutions. 

Since the pressure is constant across the boundary layer, the jet pressure in Equation \eqref{bernofN} is ``known'' from the properties of the ambient medium. Similarly, we can consider the Eddington ratio as a known or at least tunable parameter. Equation \eqref{bernofN} can therefore be considered a relationship between the jet Lorentz factor, $\Gamma_{\rm j}$, and the characteristic density of scatterers within the outflow, $\rho'_{\rm j}$. 

The comoving density of the self-similar solutions increases rapidly from the central axis of the jet. However, we physically demand that the density transition to its ambient value once the flow becomes non-relativistic and dominated by the rest-mass density of scatterers (which happens roughly simultaneously as shown in Figure \ref{fig:gammarho}). Returning to our definition of the Lorentz factor in terms of the self-similar function $f$, the flow will become non-relativistic and break self-similarity once the function $f$ satisfies $\Gamma_{\rm j} \left(r/R\right)^{m} df/d\xi \equiv \Gamma_{\rm j}(r)df/d\xi \simeq 1$, where we defined $\Gamma_{\rm j}(r) \equiv \Gamma_{\rm j}\left(r/R\right)^{m}$ for ease of notation. Using our analytic solution for $f$ in this relation and rearranging then shows that the flow becomes non-relativistic at $\xi_{\rm nr}$, where

\begin{equation}
\xi_{\rm nr} \simeq \frac{4}{9\beta}\left(\Gamma_{\rm j}(r)-1\right). \label{xinr}
\end{equation}
When the flow is highly relativistic with $\Gamma_{\rm j}(r) \gg 1$, $\xi_{\rm nr}$ is much greater than one, and correspondingly the self-similar solutions hold out to large $\xi$. In general, this expression will also depend on $r$, meaning that the transition to the non-relativistic regime occurs non-self-similarly. However, when $m = 0$ and the Lorentz factor does not vary with distance along the axis, the radial dependence in Equation \eqref{xinr} drops out and the transition to non-relativistic speeds does not destroy the self-similarity of the problem. In Appendix \ref{sec:trans} we show that when $n = q+1$ and trans-relativistic solutions can be recovered, the exact location at which the flow satisfies $\Gamma v = 0$ is very nearly linear with $\Gamma_{\rm j}$. 

Even though the above arguments suggest the transition to non-relativistic velocities and the ambient gas occurs at a value of $\xi$ that depends linearly on $\Gamma_{\rm j}(r)$, the exponential relationship between the angle $\theta$ and the self-similar variable implies that the physical, angular extent of the outflow depends only \emph{logarithmically} on the Lorentz factor along the axis. In particular, returning to the above expression for $\theta$ in terms of $\xi$ (Equation \ref{thetaofxi}) and setting $\xi = \xi_{\rm nr}$ shows that the angle $\theta_{\rm nr}$ at which the flow becomes non-relativistic is

\begin{equation}
\theta_{\rm nr} \simeq \delta\theta(r)\sqrt{\frac{2}{\beta}}\left(\frac{4}{9\beta}\ln \Gamma_{\rm j}(r)\right)^{\beta/2},
\end{equation}
which -- since $\beta \le 1$ -- shows that the ratio $\theta_{\rm nr}/\delta\theta$ scales extremely weakly with the jet Lorentz factor (and its additional radial scaling in the general case where $m \neq 0$). Thus the overall extent of the jetted region of the outflow is effectively $\delta\theta$, even though the jet Lorentz factor can change by orders of magnitude along its axis.

\subsubsection{Mass Flux and Entrainment}
\label{sec:mass}
The mass flux through the solid angle of the jet out to some self-similar angle $\xi$ is given by

\begin{equation}
\dot{M}(\xi) = 2\pi \int_0^{\xi} \rho'\Gamma v_{\rm r}r^2d\left(\frac{1}{2}\theta^2\right) = \frac{2\pi r c}{\kappa}f(\xi), \label{Mdot}
\end{equation}
where, as for the energy flux, we introduced the factor of $c$ for clarity. We see that the mass flux in the jet increases in radius and angle, owing to the fact that $f(\xi) \simeq \ln \xi$ at large $\xi$ (see Equation \ref{exsols}).

The increase in the mass flux with radius arises from two distinct effects, the first being that the cross-sectional area of the jet expands as we move farther down the jet axis, and hence geometrically we encompass more material as we recede to larger distances from the origin. The second effect comes from the angular component of the velocity, which from the definition in terms of the stream function above (Equation \ref{vtheta}) is given by 

\begin{equation}
v_{\theta} = -\delta\theta(r)\frac{1}{\sqrt{2\beta}}\left(2-\beta-\frac{n}{2}\right)f^{\beta/2}. \label{vtheta2}
\end{equation}
As expected, this component of the velocity is on the order of the boundary layer thickness, and is also negative if $2 - \beta-n/2 > 0$. This demonstrates that when the flow is collimated, so that $q \le n/2$, the $\theta$-component of the velocity points toward the axis of the jet and material is drawn inward from the ambient medium provided that $n < 2$. This entrainment serves to further augment the number of scatterers within the outflow, such that when the jet remains collimated and the cross-sectional area increases relatively slowly, the additional influx of scatterers is provided by the transverse component of the velocity that ``funnels'' material into the jet.

The increase in the mass flux in angle is due to the fact that the density is largest at large $\xi$ and that the Lorentz factor declines only as $\propto 1/\xi$. If we use the fact that $f \propto \ln \xi$ at large angles and $\xi_{\rm nr} \propto \Gamma_{\rm j}(r)$ (see Equation \ref{xinr}) as the upper limit to the integral in Equation \eqref{Mdot}, then an approximation for the total mass flux is

\begin{equation}
\dot{M} \simeq \frac{8\pi c r}{9\beta \kappa}\ln \Gamma_{\rm j}(r), \label{Mdottot}
\end{equation}
which depends only logarithmically (i.e., quite weakly) on the jet Lorentz factor. This expression is also independent of the density of scatterers at the base of the outflow, $\rho'_{\rm j}$, which arises from the fact that the cross-sectional area of the jet varies as the square of the photon mean-free path, which is just inversely proportional to the density. Thus the linear increase in the mass flux precisely balances the decline in the boundary layer thickness that accompanies the increase in $\rho'_{\rm j}$.

\section{Implications for Tidal Disruption Events and Gamma-ray Bursts}
\label{sec:implications}
Here we describe the implications of the radiation-viscous jet solutions derived in the previous section in the context of jetted TDEs and GRBs.

\subsection{Jetted tidal disruption events}
A TDE occurs when a star is ripped apart by the gravitational field of a supermassive black hole (SMBH), with roughly half of the stellar debris returning to the SMBH and forming an accretion flow (e.g., \citealt{lacy82, rees88}). The fallback rate, or the rate at which material returns to pericenter following the disruption, scales roughly as $\propto t^{-5/3}$ at late times \citep{phinney89} and can be super-Eddington (for standard radiative efficiencies on the order of 0.1) for months to years for SMBHs with masses\footnote{These estimates assume that the disruption is full and that the disrupted star is solar-like. When the disruption is partial, the fraction of accreted material can be much less \citep{guillochon13}, and the late-time falloff rate is $\propto t^{-9/4}$ \citep{coughlin19b, miles20}, implying that even if super-Eddington accretion rates were reached they would not be maintained for as long.} $\lesssim 10^{7}M_{\odot}$ \citep{evans89, wu18}. If the disrupted debris circularizes efficiently and the accretion rate tracks the fallback rate, which observations for at least UV-bright TDEs suggest is the case \citep{mockler19}, then the corresponding supercritical accretion luminosity of the hole is expected to drive winds and jets \citep{strubbe09, giannios11, coughlin14}.

So far, three events have been detected that have been argued to be jetted TDEs (though see \citealt{quataert12, woosley12}), being \emph{Swift} J1644+57 \citep{bloom11, burrows11, levan11, zauderer11}, \emph{Swift} J2058+05 \citep{cenko12}, and \emph{Swift} J1112-82 \citep{brown15}. The jetted TDE that was longest-lived and with the most complete, multi-band coverage is J6144, which was X-ray bright for $\gtrsim 500$ days (e.g., \citealt{pasham15}) and to this day has detectable radio emission (e.g., \citealt{alexander20}). In contrast, J2058 displayed jetted activity for roughly $\sim 100$ days \citep{pasham15}, while J1112 was X-ray bright for $\sim 30$ days \citep{brown15}. In all of these cases, the accretion rate onto the SMBH was almost certainly super-Eddington. Because it was the most well documented in terms of its evolution and was the longest lived, we will mainly focus on the features of J1644 in the context of our model, but the overall trends should be applicable to other jetted TDEs as well (both those previously observed and those detected in the future).

During the super-Eddington phase, the disc around the SMBH formed from a TDE is likely both geometrically and optically thick owing to increased pressure support from both gas and radiation. \citet{coughlin14b} suggested that, owing to the low binding energy and specific angular momentum of the returning material to the SMBH, this disc conforms to a quasi-spherical envelope with a centrifugally supported funnel; the accretion energy liberated during the accretion process is predominantly exhausted through this funnel in the form of a jet. Simulations (e.g., \citealt{sadowski16}) have confirmed that this envelope structure is approximately upheld.

The self-similar solutions for the envelope, known as zero-Bernoulli accretion (ZEBRA) flows, possess pressure profiles with radial power-law indices $n$ that vary as functions of time in a way that depends on the SMBH mass and the fallback rate. However, the qualitative trend is that $n$ adopts values that are between $1.5$ and $2.5$, with larger values achieved at earlier times and for smaller black hole masses (see Figure 3 of \citealt{wu18}). The timescale over which the power-law index varies is also comparable to the fallback time, being months to years for typical parameters, and hence $n$ can be approximated as a constant over many dynamical times near the black hole. The jet solutions should therefore apply in a quasi-time-steady manner, in that we can treat the pressure power-law index -- and thus the jet properties -- as a sequence of steady-state self-similar solutions. The Eddington factor is also time-dependent on similar timescales as the pressure power-law index, and reaches maximum values between $1000$ and $\sim 1$ as the black hole mass is in the range $10^5$ to $10^7$ (see Figure 5 of \citealt{wu18}). The overall radial extent of the envelope is set by the location outside of which photons are no longer trapped in the flow, which yields a characteristic envelope radius on the order of $\sim few \times 10^{14}$ cm (Figure 6 of \citealt{wu18})

Near the gravitational radius of the SMBH, the ZEBRA solutions have $p'_{\rm a} \simeq \rho'_{\rm a}c^2$ for the normalization of the ambient pressure $p'_{\rm a}$ in terms of the ambient density $\rho'_{\rm a}$. We thus expect that, for jetted TDEs, the relationship between the Eddington factor and the Bernoulli parameter that results from energy conservation (Equation \ref{bernofN}) implies 

\begin{equation}
\Gamma_{\rm j}\frac{\rho'_{\rm a}}{\rho'_{\rm j}} \simeq 10-100. \label{Gammatde}
\end{equation}
The self-similar solutions do not yield a straightforward way of estimating the density of scatterers within the outflow, but we expect in general that $\rho'_{\rm j} \lesssim \rho'_{\rm a}$. Equation \eqref{Gammatde} therefore suggests that the jets from TDEs at the base of the outflow are not highly relativistic, and possess Lorentz factors that are likely on the order of tens at most; this result is consistent with the observations of \emph{Swift} J1644+57, the inferred Lorentz factors for which were on the order of 5-10 \citep{zauderer11}.

The ambient density profile of the ZEBRA solution has a power-law decline with the power-law index $q = n-1$. It is not necessarily the case that the density profile within the jetted outflow maintains the same radial dependence as that of the ambient medium; however, it seems likely that if the jet draws in material from the ambient environment, as is the case when the jet is collimated and the pressure profile falls off with radius less steeply than $\propto r^{-2}$ (see Section \ref{sec:mass}), then the power-law decline of the density within the jet should be close to that of the ambient gas. When the radial power-law index of the density and pressure are related via $q = n-1$, the jet Lorentz factor is a constant, which suggests that there is not much variation between the Lorentz factor at the base of the outflow (Equation \ref{Gammatde}) and when it reaches the edge of the envelope. 

At early times, the power-law index of the pressure profile from the ZEBRA solutions satisfies $n > 2$; as we saw above (see Section \ref{sec:collimation}), however, for these relatively steep pressure profiles the jet loses causal connectedness and expands supersonically. In this regime, therefore, we might expect the outflow to form but then shock against the natal disc, resulting in additional recollimation at the expense of blowing off part of the envelope as the shock advances through it. We thus suggest that one might expect more variability in the emission features of the outflow at early times, induced by the jet repeatedly shocking against the surrounding envelope that is continuously blown apart and reforms, which persists until the pressure profile attains a power-law index that satisfies $n \le 2$. This picture, while clearly drawn only at the qualitative level, is at least consistent with the early ($\lesssim 1$ week from detection), highly time variable behavior that was displayed by \emph{Swift} J1644+57 (e.g., \citealt{saxton12}). 

From the structured nature of the jet, most of the mass in the outflow is contained at large angles from the axis and moves with reduced Lorentz factors. Thus, if the radio emission from \emph{Swift} J1644+57 is generated by synchrotron emission as material within the outflow shocks against the surrounding medium (which observations suggest is the case; e.g., \citealt{eftekhari18, alexander20}), we would predict the majority of the radio to come from wider angles where the fluid is less relativistic. We therefore expect the X-ray emission, which originates from deeper within the outflow and can be seen along lines of sight that are aligned or nearly aligned with the axis of the jet, and the radio emission to arise from two distinct components that possess different Lorentz factors. Interestingly, this is precisely one mechanism by which the distinct, temporal peaks in the radio and X-ray emission of \emph{Swift} J1644+57 can be explained (e.g., \citealt{wang14, mimica15}). 

\subsection{Gamma-ray bursts}
\subsubsection{Long gamma-ray bursts}
Long gamma-ray bursts (GRBs), with durations $\gtrsim 1-2$ s, are thought to arise from the collapse of rapidly rotating, massive stars (the collapsar model; \citealt{woosley93, macfadyen99}), in which the radiation generated during the formation and evolution of an accretion flow around the natal black hole or neutron star gives rise to an outflow (or ``fireball'' composed of particles and radiation; e.g., \citealt{meszaros92,rees92}) that punches its way through the overlying stellar envelope. As the jet propagates through the envelope, it generates a forward shock-contact discontinuity-reverse shock structure -- the ``head'' of the jet (e.g., \citealt{begelman89, matzner03}) -- that is relatively confined in space. The breakout of the jet from the envelope produces the prompt burst of gamma-rays, though the precise mechanism for producing the radiation is still debated, i.e., whether the shocks internal to and external to the outflow can explain the emission \citep{meszaros97} or whether there is a significant, boosted and thermal component that arises from the outflow itself (e.g., \citealt{goodman86, rees05}). 

For hydrogen and helium-depleted Wolf-Rayet progenitors, which give rise to type Ibc supernovae that have been associated with long GRBs (e.g., \citealt{woosley06} and references therein), the accretion rates that result from pure freefall onto the core are comparable to $\sim 1 M_{\odot}$ s$^{-1}$ (e.g., Figure 10 of \citealt{fernandez18}). For a black hole mass of $\sim 5 M_{\odot}$, the radiative efficiency associated with the accretion process would have to be $\lesssim 10^{-10}$ in order for the luminosity to be below the Eddington limit of the hole. Therefore, the jet Lorentz factor at the scale radius (presumed to coincide, modulo a factor of a few, with the gravitational radius of the black hole or the surface of the neutron star) given in Equation \eqref{bernofN} can easily be in excess of tens to hundreds of thousands depending on how small the rest-mass density of scatterers is within the flow. 

Near the base of the outflow, it is plausible that physical conditions at angles outside of the jetted region are qualitatively similar to those described in the ZEBRA model of \citet{coughlin14b}, i.e., a highly optically and geometrically thick structure that mediates accretion onto the compact object from the infalling and surrounding stellar envelope, or a more general (but self-similar) advection-dominated accretion flow \citep{narayan94, narayan95} or adiabatic inflow-outflow solution \citep{blandford99, blandford04}. As such, it is likely that near the base of the outflow, the Lorentz factor does not vary too dramatically with radius (as above, assuming and arguing that the power-law index of the density profile is related to that of the pressure via $q = n-1$). However, eventually the jet becomes confined by the overlying stellar envelope or the ``cocoon'' of spent material that is ejected sideways at the contact discontinuity (e.g., \citealt{begelman89, bromberg18}). The pressure within the cocoon is likely to be roughly constant or at least weakly dependent on radius within the star, as the sound speed within the cocoon is likely to be larger than the speed at which the cocoon propagates into the stellar envelope (similar to the way in which the rapid increase in the post-shock sound speed of the Sedov-Taylor blastwave results in a roughly constant interior pressure; \citealt{taylor50, sedov59}). In this case, therefore, the pressure profile no longer aids in driving and sustaining the Lorentz factor of the outflow, and instead we expect the Lorentz factor of the jet to decline according to

\begin{equation}
\Gamma_{\rm j}(r) \propto r^{-\left(1+q\right)}.
\end{equation}
Thus, in the scenario when the density of scatterers is a constant (which is likely to be the case since the density in the cocoon will be weakly varying with radius and the collimation of the outflow implies lateral entrainment), we expect the Lorentz factor to decline roughly by the factor $\left(R_{\rm cc}/R_{\rm disc}\right)^{-1}$ by the time it reaches the stellar surface, where $R_{\rm cc}$ is the radius of the progenitor at core collapse and $R_{\rm disc}$ is the radius of the inner accretion flow. It is outside the scope of the present paper to estimate the radius of the disc and thus the factor by which the jet Lorentz factor would decelerate by the time it penetrates the surface; however, at the very least this fairly rapid deceleration points to the conclusion, as also argued by \citet{matzner03}, that the progenitors of long GRBs must be relatively compact to yield a highly relativistic jet under the collapsar paradigm.

Our model does not directly incorporate any dissipation through shocks, which, as discussed above, is thought to be one mechanism by which the gamma-rays are produced in GRBs (although one could calculate the dissipation that would follow as a consequence of the structured jets proposed here encountering an external medium of a given density profile). Instead, if the optical depth along the line of sight is sufficiently low, i.e., if the circumstellar environment is sufficiently ``clean'' to allow viewing of the photons emitted from the jetted region itself, the structured jets here yield an effective photosphere from which radiation is emitted that contributes to the spectrum of a gamma-ray burst. Owing to the large optical depth within the comoving frame of a given fluid element, such a spectrum would be effectively thermal, but would be a composite of all the regions that satisfy the $\tau \sim 1$ constraint necessary to reach an observer at a certain viewing angle. This radiation would then contribute to the thermal radiation -- taken over the whole emitting surface -- of the spectrum of a gamma-ray burst (e.g., \citealt{rees05}). 

There is also the relatively newly discovered class of ``ultra-long'' GRBs, which have detectable gamma-ray emission from thousands to tens of thousands of seconds following the prompt burst \citep{levan14}. The origin of these systems and the source of the energy injection is still debated, with progenitors including the collapse of blue supergiants (e.g., \citealt{perna18}) and spinning-down magnetars (e.g., \citealt{greiner15}). To the extent that a collapsar-like model also explains the origin of these systems, the structured, radiation-driven jet model outlined here should also apply in a qualitatively similar way to ordinary, long GRBs. 

\subsubsection{Short gamma-ray bursts}
Though it was suggested for some time that short GRBs likely originate from the mergers of compact objects (at least one of which is baryon rich; e.g., \citealt{paczynski86, goodman86, eichler89}; see \citealt{berger14} for a review), the discovery of GW170817 \citep{abbott17} and the concurrent GRB/kilonova \citep{abbott17b, kasen17} conclusively demonstrated that this is indeed the case (for at least some fraction of short GRBs). During the merger of two compact objects, the less massive object is tidally disrupted during the final, gravitational-wave and tidal-driven plunge into the more massive object and ejects a tidal tail of ejecta that can contain up to tens of percent of the mass of the star (e.g., \citealt{rasio94, rasio95, lee99, lee07, shibata19}). A large fraction of this ejecta accretes onto the black hole, which forms a disc that at early times is hot enough to be cooled by neutrinos (e.g., \citealt{popham99}) and can launch a neutrino-driven wind (e.g., \citealt{siegel18}). During this process a collimated jet is also likely launched (though the observational evidence is more sparse than for long GRBs; see \citealt{fong13}), which could be driven by neutrino annihilation primarily along the axis of the binary (e.g., \citealt{goodman87, meszaros92b}) or magnetic effects (e.g., \citealt{rezzolla11}). 

In these circumstances, the disc forms and launches an outflow on a timescale that is comparable to the inspiral time of the binary itself, which is on the order of the light crossing time over the gravitational radius of the black hole or the radius of the neutron star. The latter is, however, also equal to the dynamical timescale near the base of the jet, and it is therefore not clear that ignoring the explicit time dependence in the fluid equations -- which underlies the steady-state nature of the jet solutions described here -- is justified. However, for short bursts that are toward the long end of the spectrum and that have durations on the order of $\sim 0.1 -1$ s, the total duration of the burst is many ($\gtrsim 10^{5}$ for a black hole with mass $2 M_{\odot}$ and $\gtrsim 10^{4}$ for a neutron star with radius 10 km) sound crossing times near the base of the outflow. In these instances the steady-state assumption is more reasonable to adopt, and we would expect that when radiation (or relativistic particles such as neutrinos or electron-positron pairs) dominates the rest mass energy of the fluid that comprises the outflow, the structured jet solutions outlined here would apply.

Structured jets have been invoked to explain the appearance of the X-ray and radio afterglow from the short GRB\footnote{See also \citet{cunningham20}, who recently applied a Gaussian jet model to the long GRB 160625B, finding that the data preferred a Gaussian jet over a tophat model.} that accompanied GW170817 (e.g., \citealt{lamb17, alexander18, lamb18, lyman18, troja18, fong19, lamb19}). Interestingly, a model for the jet structure that is consistent with the observations at multiple different wavebands is a Gaussian profile (in energy), while power-law and top-hat profiles have been ruled out \citep{troja17,troja18}. The model that we have outlined in this paper, in which the jet structure is mediated by radiation throughout the outflow, therefore gives a theoretical and physical model from which a Gaussian profile emerges naturally.

In addition to the prompt spike of gamma-rays, a number of short GRBs have been found to display ``extended emission'' -- X-ray and gamma-ray emission that lasts for hundreds of seconds following the initial burst and that is distinct from the emission generated between the outflow and the circumburst medium \citep{lazzati01, della06, perley09, burns18}. A possible origin for this component is the prolonged fallback of weakly bound tidal ejecta to the black hole or neutron star, which circularizes -- likely through general relativistic apsidal precession, owing to the relativistic pericenter of the disrupted object \citep{rosswog02} -- and accretes onto the compact object (e.g., \citealt{faber06, lee07, metzger10, desai19, coughlin20}). With this underlying physical mechanism for generating the extended emission, we might expect the accretion flow around the compact object to evolve in a way that is similar to the ZEBRA prescription for a TDE, but with an accretion rate that is much larger in terms of the Eddington luminosity of the black hole or neutron star. It therefore seems reasonable to suspect the jets from the extended emission in short GRBs to be modulated in a manner that is similar to those from TDEs in terms of the pressure profile of the surrounding envelope, but with jet Lorentz factors that are correspondingly boosted owing to the higher Eddington fraction. While it is outside the scope of the present paper to perform a more detailed analysis of this stage, it seems that a quantitative investigation along these lines could yield useful information into the nature of the disc structure and jetted outflow during this evolutionary stage of a short GRB.

\section{Summary and conclusions}
\label{sec:summary}
We have shown that self-similar, ultra-relativistic, structured jets are solutions to the equations of radiation hydrodynamics in the viscous limit (Section \ref{sec:basic}). When the energy flux through the outflow is conserved, which is physically applicable when the jet is powered by a compact energy source, the solutions for the comoving density and Lorentz factor as functions of angle and radius are analytic and able to be written in simple, closed forms (Section \ref{sec:jets}). Near the jet axis and over a characteristic angular scale of the square root of the photon mean free path, the fluid reaches a Lorentz factor above that of the surrounding outflow, and hence the system conforms to a core of fast-moving material surrounded by a sheath of slower moving fluid; the density within the core is also reduced relative to the sheath.

The self-similar approximation necessitates that the pressure and density profiles vary as power-laws in radius along the jet axis. The qualitative properties of the jet solutions change as a function of the pressure power-law index, $n$, and the density power-law index, $q$, as described in Section \ref{sec:properties}. Among the important features are that the jet is collimated, with the flow lines asymptotically closing up to the jet axis, if $q \le n/2$; the fluid accelerates to larger Lorentz factors if $n - q > 1$; and the jet remains causally connected if $n \le 2$. When the pressure and density power-law indices satisfy $q = n-1$, the jet Lorentz factor is a constant along the jet axis, and there are trans-relativistic solutions that do not require the ultra-relativistic approximation (see Appendix \ref{sec:trans}). When the pressure profile falls off more steeply than $\propto r^{-2}$, the lack of causal connectedness likely implies that the jet shocks against its ambient environment, which will introduce additional length scales and likely violate the self-similar assumption.

In addition to the restrictions imposed by self-similarity, the only other assumptions underlying this model are that the energy flux powering the outflow is conserved, so that the jet draws its power from the physical processes taking place near the origin, and that the equation of state is relativistic (rest-mass energy much less than the enthalpy) and dominated by radiation (or particles that act effectively like radiation). The most obvious application of such a model, i.e., where these physical considerations are most readily established, is to a hyperaccreting system such as a gamma-ray burst or tidal disruption event, and this model makes a number of predictions and has a number of implications for such systems (Section \ref{sec:implications}). For example, the angular profile of the Lorentz factor and energy flux density (the latter just being proportional to $\Gamma^2$ here as the comoving pressure is independent of angle) is either Gaussian or Gaussian-like depending on the pressure and density profiles, which is one of the models for the jet structure that explains the X-ray and radio emission from GW/GRB170817 \citep{troja19}. These solutions therefore provide a physical origin for a Gaussian jet. 

The structured jet solutions described here are agnostic to the mechanism responsible for launching the relativistic jet, and in particular they do not incorporate any constraints related to the way in which the outflow achieves the Lorentz factor $\Gamma_{\rm j}$ at the scale radius. As such, instead of being highly relativistic near the black hole event horizon (or the surface of the neutron star) as suggested in Section \ref{sec:implications} when considering the application of this model to jetted TDEs and short and long GRBs, it may be the case that the outflow accelerates initially in an adiabatic manner -- with the corresponding boundary layer initially confined to the angular extremities of the outflow -- and it reaches the viscous, self-similar stage at a larger distance than the one at which the energy injection takes place. In this case, the effective scale radius appropriate to the viscous evolution would be much larger, and that radius would still have to be smaller than the total, radial extent of the confining medium for these structured jet solutions to characterize the outflow. 

When $n \le 2$, the jet draws in mass from the surrounding sheath (there are also solutions in which this is the case for $n > 2$, but in this case the jet is out of internal causal contact and likely forms shocks), which serves to provide an additional reservoir of particles that maintain the moderate-to-high optical depth of the system. Such a high optical depth is necessary not only for ensuring the consistency of our approach, which treats the interactions between the scatterers and photons within the flow in the diffusive regime, but also for shielding the fast-moving core from the further decelerating effects of larger-scale radiation drag \citep{phinney82}. The jet is therefore ``parasitic'' from the standpoint that it draws its scatterers in from its surroundings. 

The lack of causal connectedness when the pressure profile falls off more steeply than $p' \propto r^{-2}$ destroys the self-consistency of the jet model outlined here, and this feature must be considered when applying these solutions to astrophysical systems. For example, while the accretion flows from tidal disruption events possess enough angular momentum to maintain a relatively shallow pressure profile, the same may not be true for short gamma-ray bursts where the tidal disruption radius of the neutron star is comparable to the gravitational radius of the denser object. Because of the small angular momentum budget, we would expect the envelope generated from the fallback in short gamma-ray bursts to be very nearly spherically symmetric initially, in which case (for a ZEBRA or ADAF-like accretion flow) the pressure profile will decline as $p' \propto r^{-4}$. Such a rapidly declining pressure profile is incapable of keeping the jet causally connected and collimated, and hence a sufficiently large amount of mass must be lost (i.e., accreted or ejected through the interaction of the jet and the natal disc) before the self-similar solutions described here can be applied.

The equations of radiation hydrodynamics in the viscous limit are derived from the Boltzmann equation through a Chapman-Enskog expansion in the photon mean free path, which permits a self-consistent (to leading order in the mean free path) determination of the energy-momentum tensor of the radiation field from moments of the photon distribution function. This approach contrasts other ``closure schemes'' that relate the off-diagonal elements of the radiation energy-momentum tensor -- which yield the anisotropic momentum fluxes in the fluid frame -- to the energy density and energy fluxes. Closure schemes are usually constructed to match the infinitely optically thick (i.e., the Eddington limit) and optically thin limits and are ad-hoc interpolations in between (e.g., \citealt{levermore81}). However, agreement in the asymptotic regimes does not necessarily imply that the closure scheme correctly captures the effects of radiation viscosity, which are first-order (in the photon mean free path) corrections to the comoving momentum fluxes owing to local aberrations in the specific intensity. For example, M1 \citep{levermore84} reduces correctly to both the Eddington and streaming limits, but yields off-diagonal elements of the radiation energy-momentum tensor that are \emph{second order} in the energy flux (see Equations 34 -- 37 of \citealt{sadowski13}). Since the energy flux of the radiation is proportional to the photon mean-free path in the optically thick limit, the viscous terms -- which are first order in the photon mean free path -- are not captured by this approximation.  Thus, the viscous transport of momentum from the core to the sheath is underrepresented by such a closure method.

\section*{Acknowledgements}
We thank the referee, Geoffrey Bicknell, for constructive comments and suggestions that improved the quality of this paper. ERC acknowledges support from NASA through the Hubble Fellowship Program, grant \#HST-HF2-51433.001-A awarded by the Space Telescope Science Institute, which is operated by the Association of Universities for Research in Astronomy, Incorporated, under NASA contract NAS5-26555, and the National Science Foundation through grant AST-2006684. MCB acknowledges support from NASA through the Astrophysics Theory Program, Grant NNX16AI40G.

\section*{Data Availability}
No new data were generated or analysed in support of this research.

\bibliographystyle{mnras}

\begin{thebibliography}{}
\makeatletter
\relax
\def\mn@urlcharsother{\let\do\@makeother \do\$\do\&\do\#\do\^\do\_\do\%\do\~}
\def\mn@doi{\begingroup\mn@urlcharsother \@ifnextchar [ {\mn@doi@}
  {\mn@doi@[]}}
\def\mn@doi@[#1]#2{\def\@tempa{#1}\ifx\@tempa\@empty \href
  {http://dx.doi.org/#2} {doi:#2}\else \href {http://dx.doi.org/#2} {#1}\fi
  \endgroup}
\def\mn@eprint#1#2{\mn@eprint@#1:#2::\@nil}
\def\mn@eprint@arXiv#1{\href {http://arxiv.org/abs/#1} {{\tt arXiv:#1}}}
\def\mn@eprint@dblp#1{\href {http://dblp.uni-trier.de/rec/bibtex/#1.xml}
  {dblp:#1}}
\def\mn@eprint@#1:#2:#3:#4\@nil{\def\@tempa {#1}\def\@tempb {#2}\def\@tempc
  {#3}\ifx \@tempc \@empty \let \@tempc \@tempb \let \@tempb \@tempa \fi \ifx
  \@tempb \@empty \def\@tempb {arXiv}\fi \@ifundefined
  {mn@eprint@\@tempb}{\@tempb:\@tempc}{\expandafter \expandafter \csname
  mn@eprint@\@tempb\endcsname \expandafter{\@tempc}}}

\bibitem[\protect\citeauthoryear{{Abbott} et~al.,}{{Abbott}
  et~al.}{2017a}]{abbott17}
{Abbott} B.~P.,  et~al., 2017a, \mn@doi [\prl]
  {10.1103/PhysRevLett.119.161101}, \href
  {https://ui.adsabs.harvard.edu/abs/2017PhRvL.119p1101A} {119, 161101}

\bibitem[\protect\citeauthoryear{{Abbott} et~al.,}{{Abbott}
  et~al.}{2017b}]{abbott17b}
{Abbott} B.~P.,  et~al., 2017b, \mn@doi [\apjl] {10.3847/2041-8213/aa91c9},
  \href {https://ui.adsabs.harvard.edu/abs/2017ApJ...848L..12A} {848, L12}

\bibitem[\protect\citeauthoryear{{Alexander} et~al.,}{{Alexander}
  et~al.}{2018}]{alexander18}
{Alexander} K.~D.,  et~al., 2018, \mn@doi [\apjl] {10.3847/2041-8213/aad637},
  \href {https://ui.adsabs.harvard.edu/abs/2018ApJ...863L..18A} {863, L18}

\bibitem[\protect\citeauthoryear{{Alexander}, {van Velzen}, {Horesh}  \&
  {Zauderer}}{{Alexander} et~al.}{2020}]{alexander20}
{Alexander} K.~D.,  {van Velzen} S.,  {Horesh} A.,   {Zauderer} B.~A.,  2020,
  \mn@doi [\ssr] {10.1007/s11214-020-00702-w}, \href
  {https://ui.adsabs.harvard.edu/abs/2020SSRv..216...81A} {216, 81}

\bibitem[\protect\citeauthoryear{{Arav} \& {Begelman}}{{Arav} \&
  {Begelman}}{1992}]{arav92}
{Arav} N.,  {Begelman} M.~C.,  1992, \mn@doi [\apj] {10.1086/172045}, \href
  {http://adsabs.harvard.edu/abs/1992ApJ...401..125A} {401, 125}

\bibitem[\protect\citeauthoryear{{Begelman} \& {Cioffi}}{{Begelman} \&
  {Cioffi}}{1989}]{begelman89}
{Begelman} M.~C.,  {Cioffi} D.~F.,  1989, \mn@doi [\apjl] {10.1086/185542},
  \href {https://ui.adsabs.harvard.edu/abs/1989ApJ...345L..21B} {345, L21}

\bibitem[\protect\citeauthoryear{{Berger}}{{Berger}}{2014}]{berger14}
{Berger} E.,  2014, \mn@doi [\araa] {10.1146/annurev-astro-081913-035926},
  \href {https://ui.adsabs.harvard.edu/abs/2014ARA&A..52...43B} {52, 43}

\bibitem[\protect\citeauthoryear{{Blandford} \& {Begelman}}{{Blandford} \&
  {Begelman}}{1999}]{blandford99}
{Blandford} R.~D.,  {Begelman} M.~C.,  1999, \mn@doi [\mnras]
  {10.1046/j.1365-8711.1999.02358.x}, \href
  {https://ui.adsabs.harvard.edu/abs/1999MNRAS.303L...1B} {303, L1}

\bibitem[\protect\citeauthoryear{{Blandford} \& {Begelman}}{{Blandford} \&
  {Begelman}}{2004}]{blandford04}
{Blandford} R.~D.,  {Begelman} M.~C.,  2004, \mn@doi [\mnras]
  {10.1111/j.1365-2966.2004.07425.x}, \href
  {https://ui.adsabs.harvard.edu/abs/2004MNRAS.349...68B} {349, 68}

\bibitem[\protect\citeauthoryear{{Blandford} \& {Znajek}}{{Blandford} \&
  {Znajek}}{1977}]{blandford77}
{Blandford} R.~D.,  {Znajek} R.~L.,  1977, \mn@doi [\mnras]
  {10.1093/mnras/179.3.433}, \href
  {https://ui.adsabs.harvard.edu/abs/1977MNRAS.179..433B} {179, 433}

\bibitem[\protect\citeauthoryear{{Bloom} et~al.,}{{Bloom}
  et~al.}{2011}]{bloom11}
{Bloom} J.~S.,  et~al., 2011, \mn@doi [Science] {10.1126/science.1207150},
  \href {https://ui.adsabs.harvard.edu/abs/2011Sci...333..203B} {333, 203}

\bibitem[\protect\citeauthoryear{{Braginskii}}{{Braginskii}}{1965}]{braginskii65}
{Braginskii} S.~I.,  1965, Reviews of Plasma Physics, \href
  {https://ui.adsabs.harvard.edu/abs/1965RvPP....1..205B} {1, 205}

\bibitem[\protect\citeauthoryear{{Bromberg} \& {Levinson}}{{Bromberg} \&
  {Levinson}}{2007}]{bromberg07}
{Bromberg} O.,  {Levinson} A.,  2007, \mn@doi [\apj] {10.1086/522668}, \href
  {https://ui.adsabs.harvard.edu/abs/2007ApJ...671..678B} {671, 678}

\bibitem[\protect\citeauthoryear{{Bromberg}, {Tchekhovskoy}, {Gottlieb},
  {Nakar}  \& {Piran}}{{Bromberg} et~al.}{2018}]{bromberg18}
{Bromberg} O.,  {Tchekhovskoy} A.,  {Gottlieb} O.,  {Nakar} E.,   {Piran} T.,
  2018, \mn@doi [\mnras] {10.1093/mnras/stx3316}, \href
  {https://ui.adsabs.harvard.edu/abs/2018MNRAS.475.2971B} {475, 2971}

\bibitem[\protect\citeauthoryear{{Brown}, {Levan}, {Stanway}, {Tanvir},
  {Cenko}, {Berger}, {Chornock}  \& {Cucchiaria}}{{Brown}
  et~al.}{2015}]{brown15}
{Brown} G.~C.,  {Levan} A.~J.,  {Stanway} E.~R.,  {Tanvir} N.~R.,  {Cenko}
  S.~B.,  {Berger} E.,  {Chornock} R.,   {Cucchiaria} A.,  2015, \mn@doi
  [\mnras] {10.1093/mnras/stv1520}, \href
  {https://ui.adsabs.harvard.edu/abs/2015MNRAS.452.4297B} {452, 4297}

\bibitem[\protect\citeauthoryear{{Burns} et~al.,}{{Burns}
  et~al.}{2018}]{burns18}
{Burns} E.,  et~al., 2018, \mn@doi [\apjl] {10.3847/2041-8213/aad813}, \href
  {https://ui.adsabs.harvard.edu/abs/2018ApJ...863L..34B} {863, L34}

\bibitem[\protect\citeauthoryear{{Burrows} et~al.,}{{Burrows}
  et~al.}{2011}]{burrows11}
{Burrows} D.~N.,  et~al., 2011, \mn@doi [\nat] {10.1038/nature10374}, \href
  {https://ui.adsabs.harvard.edu/abs/2011Natur.476..421B} {476, 421}

\bibitem[\protect\citeauthoryear{{Cenko} et~al.,}{{Cenko}
  et~al.}{2012}]{cenko12}
{Cenko} S.~B.,  et~al., 2012, \mn@doi [\apj] {10.1088/0004-637X/753/1/77},
  \href {https://ui.adsabs.harvard.edu/abs/2012ApJ...753...77C} {753, 77}

\bibitem[\protect\citeauthoryear{{Chandra}, {Gammie}, {Foucart}  \&
  {Quataert}}{{Chandra} et~al.}{2015}]{chandra15}
{Chandra} M.,  {Gammie} C.~F.,  {Foucart} F.,   {Quataert} E.,  2015, \mn@doi
  [\apj] {10.1088/0004-637X/810/2/162}, \href
  {https://ui.adsabs.harvard.edu/abs/2015ApJ...810..162C} {810, 162}

\bibitem[\protect\citeauthoryear{{Coughlin} \& {Begelman}}{{Coughlin} \&
  {Begelman}}{2014a}]{coughlin14b}
{Coughlin} E.~R.,  {Begelman} M.~C.,  2014a, \mn@doi [\apj]
  {10.1088/0004-637X/781/2/82}, \href
  {https://ui.adsabs.harvard.edu/abs/2014ApJ...781...82C} {781, 82}

\bibitem[\protect\citeauthoryear{{Coughlin} \& {Begelman}}{{Coughlin} \&
  {Begelman}}{2014b}]{coughlin14}
{Coughlin} E.~R.,  {Begelman} M.~C.,  2014b, \mn@doi [\apj]
  {10.1088/0004-637X/797/2/103}, \href
  {http://adsabs.harvard.edu/abs/2014ApJ...797..103C} {797, 103}

\bibitem[\protect\citeauthoryear{{Coughlin} \& {Begelman}}{{Coughlin} \&
  {Begelman}}{2015a}]{coughlin15a}
{Coughlin} E.~R.,  {Begelman} M.~C.,  2015a, \mn@doi [\apj]
  {10.1088/0004-637X/809/1/1}, \href
  {http://adsabs.harvard.edu/abs/2015ApJ...809....1C} {809, 1}

\bibitem[\protect\citeauthoryear{{Coughlin} \& {Begelman}}{{Coughlin} \&
  {Begelman}}{2015b}]{coughlin15b}
{Coughlin} E.~R.,  {Begelman} M.~C.,  2015b, \mn@doi [\apj]
  {10.1088/0004-637X/809/1/2}, \href
  {http://adsabs.harvard.edu/abs/2015ApJ...809....2C} {809, 2}

\bibitem[\protect\citeauthoryear{Coughlin \& Nixon}{Coughlin \&
  Nixon}{2019}]{coughlin19b}
Coughlin E.~R.,  Nixon C.~J.,  2019, \mn@doi [The Astrophysical Journal]
  {10.3847/2041-8213/ab412d}, 883, L17

\bibitem[\protect\citeauthoryear{{Coughlin}, {Ro}  \& {Quataert}}{{Coughlin}
  et~al.}{2019}]{coughlin19}
{Coughlin} E.~R.,  {Ro} S.,   {Quataert} E.,  2019, \mn@doi [\apj]
  {10.3847/1538-4357/ab09ec}, \href
  {https://ui.adsabs.harvard.edu/abs/2019ApJ...874...58C} {874, 58}

\bibitem[\protect\citeauthoryear{{Coughlin}, {Nixon}, {Barnes}, {Metzger}  \&
  {Margutti}}{{Coughlin} et~al.}{2020}]{coughlin20}
{Coughlin} E.~R.,  {Nixon} C.~J.,  {Barnes} J.,  {Metzger} B.~D.,   {Margutti}
  R.,  2020, \mn@doi [\apjl] {10.3847/2041-8213/ab9a4e}, \href
  {https://ui.adsabs.harvard.edu/abs/2020ApJ...896L..38C} {896, L38}

\bibitem[\protect\citeauthoryear{{Cunningham} et~al.,}{{Cunningham}
  et~al.}{2020}]{cunningham20}
{Cunningham} V.,  et~al., 2020, arXiv e-prints, \href
  {https://ui.adsabs.harvard.edu/abs/2020arXiv200900579C} {p. arXiv:2009.00579}

\bibitem[\protect\citeauthoryear{{Della Valle} et~al.,}{{Della Valle}
  et~al.}{2006}]{della06}
{Della Valle} M.,  et~al., 2006, \mn@doi [\nat] {10.1038/nature05374}, \href
  {https://ui.adsabs.harvard.edu/abs/2006Natur.444.1050D} {444, 1050}

\bibitem[\protect\citeauthoryear{{Desai}, {Metzger}  \& {Foucart}}{{Desai}
  et~al.}{2019}]{desai19}
{Desai} D.,  {Metzger} B.~D.,   {Foucart} F.,  2019, \mn@doi [\mnras]
  {10.1093/mnras/stz644}, \href
  {https://ui.adsabs.harvard.edu/abs/2019MNRAS.485.4404D} {485, 4404}

\bibitem[\protect\citeauthoryear{{Eckart}}{{Eckart}}{1940}]{eckart40}
{Eckart} C.,  1940, \mn@doi [Physical Review] {10.1103/PhysRev.58.919}, \href
  {http://adsabs.harvard.edu/abs/1940PhRv...58..919E} {58, 919}

\bibitem[\protect\citeauthoryear{{Eftekhari}, {Berger}, {Zauderer}, {Margutti}
  \& {Alexander}}{{Eftekhari} et~al.}{2018}]{eftekhari18}
{Eftekhari} T.,  {Berger} E.,  {Zauderer} B.~A.,  {Margutti} R.,   {Alexander}
  K.~D.,  2018, \mn@doi [\apj] {10.3847/1538-4357/aaa8e0}, \href
  {https://ui.adsabs.harvard.edu/abs/2018ApJ...854...86E} {854, 86}

\bibitem[\protect\citeauthoryear{{Eichler}, {Livio}, {Piran}  \&
  {Schramm}}{{Eichler} et~al.}{1989}]{eichler89}
{Eichler} D.,  {Livio} M.,  {Piran} T.,   {Schramm} D.~N.,  1989, \mn@doi
  [\nat] {10.1038/340126a0}, \href
  {https://ui.adsabs.harvard.edu/abs/1989Natur.340..126E} {340, 126}

\bibitem[\protect\citeauthoryear{{Evans} \& {Kochanek}}{{Evans} \&
  {Kochanek}}{1989}]{evans89}
{Evans} C.~R.,  {Kochanek} C.~S.,  1989, \mn@doi [\apjl] {10.1086/185567},
  \href {https://ui.adsabs.harvard.edu/abs/1989ApJ...346L..13E} {346, L13}

\bibitem[\protect\citeauthoryear{{Faber}, {Baumgarte}, {Shapiro}, {Taniguchi}
  \& {Rasio}}{{Faber} et~al.}{2006}]{faber06}
{Faber} J.~A.,  {Baumgarte} T.~W.,  {Shapiro} S.~L.,  {Taniguchi} K.,   {Rasio}
  F.~A.,  2006, \mn@doi [\prd] {10.1103/PhysRevD.73.024012}, \href
  {https://ui.adsabs.harvard.edu/abs/2006PhRvD..73b4012F} {73, 024012}

\bibitem[\protect\citeauthoryear{{Fern{\'a}ndez}, {Quataert}, {Kashiyama}  \&
  {Coughlin}}{{Fern{\'a}ndez} et~al.}{2018}]{fernandez18}
{Fern{\'a}ndez} R.,  {Quataert} E.,  {Kashiyama} K.,   {Coughlin} E.~R.,  2018,
  \mn@doi [\mnras] {10.1093/mnras/sty306}, \href
  {https://ui.adsabs.harvard.edu/abs/2018MNRAS.476.2366F} {476, 2366}

\bibitem[\protect\citeauthoryear{{Finazzo}, {Critelli}, {Rougemont}  \&
  {Noronha}}{{Finazzo} et~al.}{2016}]{finazzo16}
{Finazzo} S.~I.,  {Critelli} R.,  {Rougemont} R.,   {Noronha} J.,  2016,
  \mn@doi [\prd] {10.1103/PhysRevD.94.054020}, \href
  {https://ui.adsabs.harvard.edu/abs/2016PhRvD..94e4020F} {94, 054020}

\bibitem[\protect\citeauthoryear{{Fong} \& {Berger}}{{Fong} \&
  {Berger}}{2013}]{fong13}
{Fong} W.,  {Berger} E.,  2013, \mn@doi [\apj] {10.1088/0004-637X/776/1/18},
  \href {https://ui.adsabs.harvard.edu/abs/2013ApJ...776...18F} {776, 18}

\bibitem[\protect\citeauthoryear{{Fong} et~al.,}{{Fong} et~al.}{2019}]{fong19}
{Fong} W.,  et~al., 2019, \mn@doi [\apjl] {10.3847/2041-8213/ab3d9e}, \href
  {https://ui.adsabs.harvard.edu/abs/2019ApJ...883L...1F} {883, L1}

\bibitem[\protect\citeauthoryear{{Foucart}, {Chandra}, {Gammie}  \&
  {Quataert}}{{Foucart} et~al.}{2016}]{foucart16}
{Foucart} F.,  {Chandra} M.,  {Gammie} C.~F.,   {Quataert} E.,  2016, \mn@doi
  [\mnras] {10.1093/mnras/stv2687}, \href
  {https://ui.adsabs.harvard.edu/abs/2016MNRAS.456.1332F} {456, 1332}

\bibitem[\protect\citeauthoryear{{Giannios} \& {Metzger}}{{Giannios} \&
  {Metzger}}{2011}]{giannios11}
{Giannios} D.,  {Metzger} B.~D.,  2011, \mn@doi [\mnras]
  {10.1111/j.1365-2966.2011.19188.x}, \href
  {https://ui.adsabs.harvard.edu/abs/2011MNRAS.416.2102G} {416, 2102}

\bibitem[\protect\citeauthoryear{{Goodman}}{{Goodman}}{1986}]{goodman86}
{Goodman} J.,  1986, \mn@doi [\apjl] {10.1086/184741}, \href
  {https://ui.adsabs.harvard.edu/abs/1986ApJ...308L..47G} {308, L47}

\bibitem[\protect\citeauthoryear{{Goodman}, {Dar}  \& {Nussinov}}{{Goodman}
  et~al.}{1987}]{goodman87}
{Goodman} J.,  {Dar} A.,   {Nussinov} S.,  1987, \mn@doi [\apjl]
  {10.1086/184840}, \href
  {https://ui.adsabs.harvard.edu/abs/1987ApJ...314L...7G} {314, L7}

\bibitem[\protect\citeauthoryear{{Greenberg}}{{Greenberg}}{1975}]{greenberg75}
{Greenberg} P.~J.,  1975, \mn@doi [\apj] {10.1086/153379}, \href
  {https://ui.adsabs.harvard.edu/abs/1975ApJ...195..761G} {195, 761}

\bibitem[\protect\citeauthoryear{{Greiner} et~al.,}{{Greiner}
  et~al.}{2015}]{greiner15}
{Greiner} J.,  et~al., 2015, \mn@doi [\nat] {10.1038/nature14579}, \href
  {https://ui.adsabs.harvard.edu/abs/2015Natur.523..189G} {523, 189}

\bibitem[\protect\citeauthoryear{Guillochon \& Ramirez-Ruiz}{Guillochon \&
  Ramirez-Ruiz}{2013}]{guillochon13}
Guillochon J.,  Ramirez-Ruiz E.,  2013, \mn@doi [The Astrophysical Journal]
  {10.1088/0004-637x/767/1/25}, 767, 25

\bibitem[\protect\citeauthoryear{{Harrison} et~al.,}{{Harrison}
  et~al.}{1999}]{harrison99}
{Harrison} F.~A.,  et~al., 1999, \mn@doi [\apjl] {10.1086/312282}, \href
  {https://ui.adsabs.harvard.edu/abs/1999ApJ...523L.121H} {523, L121}

\bibitem[\protect\citeauthoryear{{Kasen}, {Metzger}, {Barnes}, {Quataert}  \&
  {Ramirez-Ruiz}}{{Kasen} et~al.}{2017}]{kasen17}
{Kasen} D.,  {Metzger} B.,  {Barnes} J.,  {Quataert} E.,   {Ramirez-Ruiz} E.,
  2017, \mn@doi [\nat] {10.1038/nature24453}, \href
  {https://ui.adsabs.harvard.edu/abs/2017Natur.551...80K} {551, 80}

\bibitem[\protect\citeauthoryear{{Kathirgamaraju}, {Tchekhovskoy}, {Giannios}
  \& {Barniol Duran}}{{Kathirgamaraju} et~al.}{2019}]{kath19}
{Kathirgamaraju} A.,  {Tchekhovskoy} A.,  {Giannios} D.,   {Barniol Duran} R.,
  2019, \mn@doi [\mnras] {10.1093/mnrasl/slz012}, \href
  {https://ui.adsabs.harvard.edu/abs/2019MNRAS.484L..98K} {484, L98}

\bibitem[\protect\citeauthoryear{{Kohler} \& {Begelman}}{{Kohler} \&
  {Begelman}}{2012}]{kohler12}
{Kohler} S.,  {Begelman} M.~C.,  2012, \mn@doi [\mnras]
  {10.1111/j.1365-2966.2012.21876.x}, \href
  {https://ui.adsabs.harvard.edu/abs/2012MNRAS.426..595K} {426, 595}

\bibitem[\protect\citeauthoryear{{Kohler} \& {Begelman}}{{Kohler} \&
  {Begelman}}{2015}]{kohler15}
{Kohler} S.,  {Begelman} M.~C.,  2015, \mn@doi [\mnras]
  {10.1093/mnras/stu2135}, \href
  {https://ui.adsabs.harvard.edu/abs/2015MNRAS.446.1195K} {446, 1195}

\bibitem[\protect\citeauthoryear{{Kundu} \& {Cohen}}{{Kundu} \&
  {Cohen}}{2008}]{kundu08}
{Kundu} P.~K.,  {Cohen} I.~M.,  2008, {Fluid Mechanics: Fourth Edition}.
{Academic Press, Elsevier, Inc., London, England}

\bibitem[\protect\citeauthoryear{{Lacy}, {Townes}  \& {Hollenbach}}{{Lacy}
  et~al.}{1982}]{lacy82}
{Lacy} J.~H.,  {Townes} C.~H.,   {Hollenbach} D.~J.,  1982, \mn@doi [\apj]
  {10.1086/160402}, \href
  {https://ui.adsabs.harvard.edu/abs/1982ApJ...262..120L} {262, 120}

\bibitem[\protect\citeauthoryear{{Lamb} \& {Kobayashi}}{{Lamb} \&
  {Kobayashi}}{2017}]{lamb17}
{Lamb} G.~P.,  {Kobayashi} S.,  2017, \mn@doi [\mnras] {10.1093/mnras/stx2345},
  \href {https://ui.adsabs.harvard.edu/abs/2017MNRAS.472.4953L} {472, 4953}

\bibitem[\protect\citeauthoryear{{Lamb} \& {Kobayashi}}{{Lamb} \&
  {Kobayashi}}{2018}]{lamb18}
{Lamb} G.~P.,  {Kobayashi} S.,  2018, \mn@doi [\mnras] {10.1093/mnras/sty1108},
  \href {https://ui.adsabs.harvard.edu/abs/2018MNRAS.478..733L} {478, 733}

\bibitem[\protect\citeauthoryear{{Lamb} et~al.,}{{Lamb} et~al.}{2019}]{lamb19}
{Lamb} G.~P.,  et~al., 2019, \mn@doi [\apjl] {10.3847/2041-8213/aaf96b}, \href
  {https://ui.adsabs.harvard.edu/abs/2019ApJ...870L..15L} {870, L15}

\bibitem[\protect\citeauthoryear{{Lazzati}, {Ramirez-Ruiz}  \&
  {Ghisellini}}{{Lazzati} et~al.}{2001}]{lazzati01}
{Lazzati} D.,  {Ramirez-Ruiz} E.,   {Ghisellini} G.,  2001, \mn@doi [\aap]
  {10.1051/0004-6361:20011485}, \href
  {https://ui.adsabs.harvard.edu/abs/2001A&A...379L..39L} {379, L39}

\bibitem[\protect\citeauthoryear{{Lee} \& {Klu{\'z}niak}}{{Lee} \&
  {Klu{\'z}niak}}{1999}]{lee99}
{Lee} W.~H.,  {Klu{\'z}niak} W.,  1999, \mn@doi [\apj] {10.1086/307958}, \href
  {https://ui.adsabs.harvard.edu/abs/1999ApJ...526..178L} {526, 178}

\bibitem[\protect\citeauthoryear{{Lee} \& {Ramirez-Ruiz}}{{Lee} \&
  {Ramirez-Ruiz}}{2007}]{lee07}
{Lee} W.~H.,  {Ramirez-Ruiz} E.,  2007, \mn@doi [New Journal of Physics]
  {10.1088/1367-2630/9/1/017}, \href
  {https://ui.adsabs.harvard.edu/abs/2007NJPh....9...17L} {9, 17}

\bibitem[\protect\citeauthoryear{{Levan} et~al.,}{{Levan}
  et~al.}{2011}]{levan11}
{Levan} A.~J.,  et~al., 2011, \mn@doi [Science] {10.1126/science.1207143},
  \href {https://ui.adsabs.harvard.edu/abs/2011Sci...333..199L} {333, 199}

\bibitem[\protect\citeauthoryear{{Levan} et~al.,}{{Levan}
  et~al.}{2014}]{levan14}
{Levan} A.~J.,  et~al., 2014, \mn@doi [\apj] {10.1088/0004-637X/781/1/13},
  \href {https://ui.adsabs.harvard.edu/abs/2014ApJ...781...13L} {781, 13}

\bibitem[\protect\citeauthoryear{{Levermore}}{{Levermore}}{1984}]{levermore84}
{Levermore} C.~D.,  1984, \mn@doi [\jqsrt] {10.1016/0022-4073(84)90112-2},
  \href {https://ui.adsabs.harvard.edu/abs/1984JQSRT..31..149L} {31, 149}

\bibitem[\protect\citeauthoryear{{Levermore} \& {Pomraning}}{{Levermore} \&
  {Pomraning}}{1981}]{levermore81}
{Levermore} C.~D.,  {Pomraning} G.~C.,  1981, \mn@doi [\apj] {10.1086/159157},
  \href {https://ui.adsabs.harvard.edu/abs/1981ApJ...248..321L} {248, 321}

\bibitem[\protect\citeauthoryear{{Loeb} \& {Laor}}{{Loeb} \&
  {Laor}}{1992}]{loeb92}
{Loeb} A.,  {Laor} A.,  1992, \mn@doi [\apj] {10.1086/170857}, \href
  {http://adsabs.harvard.edu/abs/1992ApJ...384..115L} {384, 115}

\bibitem[\protect\citeauthoryear{{Lyman} et~al.,}{{Lyman}
  et~al.}{2018}]{lyman18}
{Lyman} J.~D.,  et~al., 2018, \mn@doi [Nature Astronomy]
  {10.1038/s41550-018-0511-3}, \href
  {https://ui.adsabs.harvard.edu/abs/2018NatAs...2..751L} {2, 751}

\bibitem[\protect\citeauthoryear{{MacFadyen} \& {Woosley}}{{MacFadyen} \&
  {Woosley}}{1999}]{macfadyen99}
{MacFadyen} A.~I.,  {Woosley} S.~E.,  1999, \mn@doi [\apj] {10.1086/307790},
  \href {https://ui.adsabs.harvard.edu/abs/1999ApJ...524..262M} {524, 262}

\bibitem[\protect\citeauthoryear{{Matzner}}{{Matzner}}{2003}]{matzner03}
{Matzner} C.~D.,  2003, \mn@doi [\mnras] {10.1046/j.1365-8711.2003.06969.x},
  \href {https://ui.adsabs.harvard.edu/abs/2003MNRAS.345..575M} {345, 575}

\bibitem[\protect\citeauthoryear{{Meszaros} \& {Rees}}{{Meszaros} \&
  {Rees}}{1992a}]{meszaros92}
{Meszaros} P.,  {Rees} M.~J.,  1992a, \mn@doi [\mnras]
  {10.1093/mnras/257.1.29P}, \href
  {https://ui.adsabs.harvard.edu/abs/1992MNRAS.257P..29M} {257, 29P}

\bibitem[\protect\citeauthoryear{{Meszaros} \& {Rees}}{{Meszaros} \&
  {Rees}}{1992b}]{meszaros92b}
{Meszaros} P.,  {Rees} M.~J.,  1992b, \mn@doi [\apj] {10.1086/171813}, \href
  {https://ui.adsabs.harvard.edu/abs/1992ApJ...397..570M} {397, 570}

\bibitem[\protect\citeauthoryear{{M{\'e}sz{\'a}ros} \&
  {Rees}}{{M{\'e}sz{\'a}ros} \& {Rees}}{1997}]{meszaros97}
{M{\'e}sz{\'a}ros} P.,  {Rees} M.~J.,  1997, \mn@doi [\apj] {10.1086/303625},
  \href {https://ui.adsabs.harvard.edu/abs/1997ApJ...476..232M} {476, 232}

\bibitem[\protect\citeauthoryear{{Metzger} \& {Stone}}{{Metzger} \&
  {Stone}}{2016}]{metzger16}
{Metzger} B.~D.,  {Stone} N.~C.,  2016, \mn@doi [\mnras]
  {10.1093/mnras/stw1394}, \href
  {https://ui.adsabs.harvard.edu/abs/2016MNRAS.461..948M} {461, 948}

\bibitem[\protect\citeauthoryear{{Metzger}, {Arcones}, {Quataert}  \&
  {Mart{\'\i}nez-Pinedo}}{{Metzger} et~al.}{2010}]{metzger10}
{Metzger} B.~D.,  {Arcones} A.,  {Quataert} E.,   {Mart{\'\i}nez-Pinedo} G.,
  2010, \mn@doi [\mnras] {10.1111/j.1365-2966.2009.16107.x}, \href
  {https://ui.adsabs.harvard.edu/abs/2010MNRAS.402.2771M} {402, 2771}

\bibitem[\protect\citeauthoryear{{Mihalas} \& {Mihalas}}{{Mihalas} \&
  {Mihalas}}{1984}]{mihalas84}
{Mihalas} D.,  {Mihalas} B.~W.,  1984, {Foundations of radiation
  hydrodynamics}.
{New York: Oxford University Press}

\bibitem[\protect\citeauthoryear{{Miles}, {Coughlin}  \& {Nixon}}{{Miles}
  et~al.}{2020}]{miles20}
{Miles} P.~R.,  {Coughlin} E.~R.,   {Nixon} C.~J.,  2020, arXiv e-prints, \href
  {https://ui.adsabs.harvard.edu/abs/2020arXiv200609375M} {p. arXiv:2006.09375}

\bibitem[\protect\citeauthoryear{{Mimica}, {Giannios}, {Metzger}  \&
  {Aloy}}{{Mimica} et~al.}{2015}]{mimica15}
{Mimica} P.,  {Giannios} D.,  {Metzger} B.~D.,   {Aloy} M.~A.,  2015, \mn@doi
  [\mnras] {10.1093/mnras/stv825}, \href
  {https://ui.adsabs.harvard.edu/abs/2015MNRAS.450.2824M} {450, 2824}

\bibitem[\protect\citeauthoryear{{Mockler}, {Guillochon}  \&
  {Ramirez-Ruiz}}{{Mockler} et~al.}{2019}]{mockler19}
{Mockler} B.,  {Guillochon} J.,   {Ramirez-Ruiz} E.,  2019, \mn@doi [\apj]
  {10.3847/1538-4357/ab010f}, \href
  {https://ui.adsabs.harvard.edu/abs/2019ApJ...872..151M} {872, 151}

\bibitem[\protect\citeauthoryear{{Narayan} \& {Yi}}{{Narayan} \&
  {Yi}}{1994}]{narayan94}
{Narayan} R.,  {Yi} I.,  1994, \mn@doi [\apjl] {10.1086/187381}, \href
  {https://ui.adsabs.harvard.edu/abs/1994ApJ...428L..13N} {428, L13}

\bibitem[\protect\citeauthoryear{{Narayan} \& {Yi}}{{Narayan} \&
  {Yi}}{1995}]{narayan95}
{Narayan} R.,  {Yi} I.,  1995, \mn@doi [\apj] {10.3847/1538-4357/ab09ec}, \href
  {https://ui.adsabs.harvard.edu/abs/2019ApJ...874...58C} {444, 231}

\bibitem[\protect\citeauthoryear{{Paczynski}}{{Paczynski}}{1986}]{paczynski86}
{Paczynski} B.,  1986, \mn@doi [\apjl] {10.1086/184740}, \href
  {https://ui.adsabs.harvard.edu/abs/1986ApJ...308L..43P} {308, L43}

\bibitem[\protect\citeauthoryear{{Panaitescu} \& {Kumar}}{{Panaitescu} \&
  {Kumar}}{2002}]{panaitescu02}
{Panaitescu} A.,  {Kumar} P.,  2002, \mn@doi [\apj] {10.1086/340094}, \href
  {https://ui.adsabs.harvard.edu/abs/2002ApJ...571..779P} {571, 779}

\bibitem[\protect\citeauthoryear{{Pasham} et~al.,}{{Pasham}
  et~al.}{2015}]{pasham15}
{Pasham} D.~R.,  et~al., 2015, \mn@doi [\apj] {10.1088/0004-637X/805/1/68},
  \href {https://ui.adsabs.harvard.edu/abs/2015ApJ...805...68P} {805, 68}

\bibitem[\protect\citeauthoryear{{Perley} et~al.,}{{Perley}
  et~al.}{2009}]{perley09}
{Perley} D.~A.,  et~al., 2009, \mn@doi [\apj] {10.1088/0004-637X/696/2/1871},
  \href {https://ui.adsabs.harvard.edu/abs/2009ApJ...696.1871P} {696, 1871}

\bibitem[\protect\citeauthoryear{{Perna}, {Lazzati}  \& {Cantiello}}{{Perna}
  et~al.}{2018}]{perna18}
{Perna} R.,  {Lazzati} D.,   {Cantiello} M.,  2018, \mn@doi [\apj]
  {10.3847/1538-4357/aabcc1}, \href
  {https://ui.adsabs.harvard.edu/abs/2018ApJ...859...48P} {859, 48}

\bibitem[\protect\citeauthoryear{{Phinney}}{{Phinney}}{1982}]{phinney82}
{Phinney} E.~S.,  1982, \mn@doi [\mnras] {10.1093/mnras/198.4.1109}, \href
  {https://ui.adsabs.harvard.edu/abs/1982MNRAS.198.1109P} {198, 1109}

\bibitem[\protect\citeauthoryear{{Phinney}}{{Phinney}}{1989}]{phinney89}
{Phinney} E.~S.,  1989, in {Morris} M.,  ed.,  IAU Symposium Vol. 136, The
  Center of the Galaxy. p.~543

\bibitem[\protect\citeauthoryear{{Popham}, {Woosley}  \& {Fryer}}{{Popham}
  et~al.}{1999}]{popham99}
{Popham} R.,  {Woosley} S.~E.,   {Fryer} C.,  1999, \mn@doi [\apj]
  {10.1086/307259}, \href
  {https://ui.adsabs.harvard.edu/abs/1999ApJ...518..356P} {518, 356}

\bibitem[\protect\citeauthoryear{{Quataert} \& {Kasen}}{{Quataert} \&
  {Kasen}}{2012}]{quataert12}
{Quataert} E.,  {Kasen} D.,  2012, \mn@doi [\mnras]
  {10.1111/j.1745-3933.2011.01151.x}, \href
  {https://ui.adsabs.harvard.edu/abs/2012MNRAS.419L...1Q} {419, L1}

\bibitem[\protect\citeauthoryear{{Rasio} \& {Shapiro}}{{Rasio} \&
  {Shapiro}}{1994}]{rasio94}
{Rasio} F.~A.,  {Shapiro} S.~L.,  1994, \mn@doi [\apj] {10.1086/174566}, \href
  {https://ui.adsabs.harvard.edu/abs/1994ApJ...432..242R} {432, 242}

\bibitem[\protect\citeauthoryear{{Rasio} \& {Shapiro}}{{Rasio} \&
  {Shapiro}}{1995}]{rasio95}
{Rasio} F.~A.,  {Shapiro} S.~L.,  1995, \mn@doi [\apj] {10.1086/175130}, \href
  {https://ui.adsabs.harvard.edu/abs/1995ApJ...438..887R} {438, 887}

\bibitem[\protect\citeauthoryear{{Rees}}{{Rees}}{1988}]{rees88}
{Rees} M.~J.,  1988, \mn@doi [\nat] {10.1038/333523a0}, \href
  {https://ui.adsabs.harvard.edu/abs/1988Natur.333..523R} {333, 523}

\bibitem[\protect\citeauthoryear{{Rees} \& {Meszaros}}{{Rees} \&
  {Meszaros}}{1992}]{rees92}
{Rees} M.~J.,  {Meszaros} P.,  1992, \mn@doi [\mnras]
  {10.1093/mnras/258.1.41P}, \href
  {https://ui.adsabs.harvard.edu/abs/1992MNRAS.258P..41R} {258, 41}

\bibitem[\protect\citeauthoryear{{Rees} \& {M{\'e}sz{\'a}ros}}{{Rees} \&
  {M{\'e}sz{\'a}ros}}{2005}]{rees05}
{Rees} M.~J.,  {M{\'e}sz{\'a}ros} P.,  2005, \mn@doi [\apj] {10.1086/430818},
  \href {https://ui.adsabs.harvard.edu/abs/2005ApJ...628..847R} {628, 847}

\bibitem[\protect\citeauthoryear{{Rezzolla}, {Giacomazzo}, {Baiotti}, {Granot},
  {Kouveliotou}  \& {Aloy}}{{Rezzolla} et~al.}{2011}]{rezzolla11}
{Rezzolla} L.,  {Giacomazzo} B.,  {Baiotti} L.,  {Granot} J.,  {Kouveliotou}
  C.,   {Aloy} M.~A.,  2011, \mn@doi [\apjl] {10.1088/2041-8205/732/1/L6},
  \href {https://ui.adsabs.harvard.edu/abs/2011ApJ...732L...6R} {732, L6}

\bibitem[\protect\citeauthoryear{{Rhoads}}{{Rhoads}}{1999}]{rhoads99}
{Rhoads} J.~E.,  1999, \mn@doi [\apj] {10.1086/307907}, \href
  {https://ui.adsabs.harvard.edu/abs/1999ApJ...525..737R} {525, 737}

\bibitem[\protect\citeauthoryear{{Rosswog} \& {Davies}}{{Rosswog} \&
  {Davies}}{2002}]{rosswog02}
{Rosswog} S.,  {Davies} M.~B.,  2002, \mn@doi [\mnras]
  {10.1046/j.1365-8711.2002.05409.x}, \href
  {https://ui.adsabs.harvard.edu/abs/2002MNRAS.334..481R} {334, 481}

\bibitem[\protect\citeauthoryear{{S{{a}}dowski}, {Narayan}, {Tchekhovskoy}  \&
  {Zhu}}{{S{{a}}dowski} et~al.}{2013}]{sadowski13}
{S{{a}}dowski} A.,  {Narayan} R.,  {Tchekhovskoy} A.,   {Zhu} Y.,  2013,
  \mn@doi [\mnras] {10.1093/mnras/sts632}, \href
  {https://ui.adsabs.harvard.edu/abs/2013MNRAS.429.3533S} {429, 3533}

\bibitem[\protect\citeauthoryear{{Sari}, {Piran}  \& {Halpern}}{{Sari}
  et~al.}{1999}]{sari99}
{Sari} R.,  {Piran} T.,   {Halpern} J.~P.,  1999, \mn@doi [\apjl]
  {10.1086/312109}, \href
  {https://ui.adsabs.harvard.edu/abs/1999ApJ...519L..17S} {519, L17}

\bibitem[\protect\citeauthoryear{{Saxton}, {Soria}, {Wu}  \& {Kuin}}{{Saxton}
  et~al.}{2012}]{saxton12}
{Saxton} C.~J.,  {Soria} R.,  {Wu} K.,   {Kuin} N. P.~M.,  2012, \mn@doi
  [\mnras] {10.1111/j.1365-2966.2012.20739.x}, \href
  {https://ui.adsabs.harvard.edu/abs/2012MNRAS.422.1625S} {422, 1625}

\bibitem[\protect\citeauthoryear{{Sedov}}{{Sedov}}{1959}]{sedov59}
{Sedov} L.~I.,  1959, {Similarity and Dimensional Methods in Mechanics}.
{Academic Press, New York, NY}

\bibitem[\protect\citeauthoryear{{Shibata} \& {Hotokezaka}}{{Shibata} \&
  {Hotokezaka}}{2019}]{shibata19}
{Shibata} M.,  {Hotokezaka} K.,  2019, \mn@doi [Annual Review of Nuclear and
  Particle Science] {10.1146/annurev-nucl-101918-023625}, \href
  {https://ui.adsabs.harvard.edu/abs/2019ARNPS..69...41S} {69, 41}

\bibitem[\protect\citeauthoryear{{Siegel} \& {Metzger}}{{Siegel} \&
  {Metzger}}{2018}]{siegel18}
{Siegel} D.~M.,  {Metzger} B.~D.,  2018, \mn@doi [\apj]
  {10.3847/1538-4357/aabaec}, \href
  {https://ui.adsabs.harvard.edu/abs/2018ApJ...858...52S} {858, 52}

\bibitem[\protect\citeauthoryear{{S{\k{a}}dowski}, {Tejeda}, {Gafton},
  {Rosswog}  \& {Abarca}}{{S{\k{a}}dowski} et~al.}{2016}]{sadowski16}
{S{\k{a}}dowski} A.,  {Tejeda} E.,  {Gafton} E.,  {Rosswog} S.,   {Abarca} D.,
  2016, \mn@doi [\mnras] {10.1093/mnras/stw589}, \href
  {https://ui.adsabs.harvard.edu/abs/2016MNRAS.458.4250S} {458, 4250}

\bibitem[\protect\citeauthoryear{{Strubbe} \& {Quataert}}{{Strubbe} \&
  {Quataert}}{2009}]{strubbe09}
{Strubbe} L.~E.,  {Quataert} E.,  2009, \mn@doi [\mnras]
  {10.1111/j.1365-2966.2009.15599.x}, \href
  {https://ui.adsabs.harvard.edu/abs/2009MNRAS.400.2070S} {400, 2070}

\bibitem[\protect\citeauthoryear{{Taylor}}{{Taylor}}{1950}]{taylor50}
{Taylor} G.,  1950, \mn@doi [Proceedings of the Royal Society of London Series
  A] {10.1098/rspa.1950.0049}, \href
  {https://ui.adsabs.harvard.edu/abs/1950RSPSA.201..159T} {201, 159}

\bibitem[\protect\citeauthoryear{{Troja} et~al.,}{{Troja}
  et~al.}{2017}]{troja17}
{Troja} E.,  et~al., 2017, \mn@doi [\nat] {10.1038/nature24290}, \href
  {https://ui.adsabs.harvard.edu/abs/2017Natur.551...71T} {551, 71}

\bibitem[\protect\citeauthoryear{{Troja} et~al.,}{{Troja}
  et~al.}{2018}]{troja18}
{Troja} E.,  et~al., 2018, \mn@doi [\mnras] {10.1093/mnrasl/sly061}, \href
  {https://ui.adsabs.harvard.edu/abs/2018MNRAS.478L..18T} {478, L18}

\bibitem[\protect\citeauthoryear{{Troja} et~al.,}{{Troja}
  et~al.}{2019}]{troja19}
{Troja} E.,  et~al., 2019, \mn@doi [\mnras] {10.1093/mnras/stz2248}, \href
  {https://ui.adsabs.harvard.edu/abs/2019MNRAS.489.1919T} {489, 1919}

\bibitem[\protect\citeauthoryear{{Wang}, {Lei}, {Wang}, {Zou}, {Zhang}, {Gao}
  \& {Huang}}{{Wang} et~al.}{2014}]{wang14}
{Wang} J.-Z.,  {Lei} W.-H.,  {Wang} D.-X.,  {Zou} Y.-C.,  {Zhang} B.,  {Gao}
  H.,   {Huang} C.-Y.,  2014, \mn@doi [\apj] {10.1088/0004-637X/788/1/32},
  \href {https://ui.adsabs.harvard.edu/abs/2014ApJ...788...32W} {788, 32}

\bibitem[\protect\citeauthoryear{{Weinberg}}{{Weinberg}}{1971}]{weinberg71}
{Weinberg} S.,  1971, \mn@doi [\apj] {10.1086/151073}, \href
  {http://adsabs.harvard.edu/abs/1971ApJ...168..175W} {168, 175}

\bibitem[\protect\citeauthoryear{{Woosley}}{{Woosley}}{1993}]{woosley93}
{Woosley} S.~E.,  1993, \mn@doi [\apj] {10.1086/172359}, \href
  {https://ui.adsabs.harvard.edu/abs/1993ApJ...405..273W} {405, 273}

\bibitem[\protect\citeauthoryear{{Woosley} \& {Bloom}}{{Woosley} \&
  {Bloom}}{2006}]{woosley06}
{Woosley} S.~E.,  {Bloom} J.~S.,  2006, \mn@doi [\araa]
  {10.1146/annurev.astro.43.072103.150558}, \href
  {https://ui.adsabs.harvard.edu/abs/2006ARA&A..44..507W} {44, 507}

\bibitem[\protect\citeauthoryear{{Woosley} \& {Heger}}{{Woosley} \&
  {Heger}}{2012}]{woosley12}
{Woosley} S.~E.,  {Heger} A.,  2012, \mn@doi [\apj]
  {10.1088/0004-637X/752/1/32}, \href
  {https://ui.adsabs.harvard.edu/abs/2012ApJ...752...32W} {752, 32}

\bibitem[\protect\citeauthoryear{{Wu}, {Coughlin}  \& {Nixon}}{{Wu}
  et~al.}{2018}]{wu18}
{Wu} S.,  {Coughlin} E.~R.,   {Nixon} C.,  2018, \mn@doi [\mnras]
  {10.1093/mnras/sty971}, \href
  {https://ui.adsabs.harvard.edu/abs/2018MNRAS.478.3016W} {478, 3016}

\bibitem[\protect\citeauthoryear{{Zauderer} et~al.,}{{Zauderer}
  et~al.}{2011}]{zauderer11}
{Zauderer} B.~A.,  et~al., 2011, \mn@doi [\nat] {10.1038/nature10366}, \href
  {https://ui.adsabs.harvard.edu/abs/2011Natur.476..425Z} {476, 425}

\bibitem[\protect\citeauthoryear{{Zrake}, {Beloborodov}  \& {Lundman}}{{Zrake}
  et~al.}{2019}]{zrake19}
{Zrake} J.,  {Beloborodov} A.~M.,   {Lundman} C.,  2019, \mn@doi [\apj]
  {10.3847/1538-4357/ab364b}, \href
  {https://ui.adsabs.harvard.edu/abs/2019ApJ...885...30Z} {885, 30}

\makeatother
\end{thebibliography}

\appendix
\section{Derivation of the Bernoulli and entropy equations}
\label{sec:appA}
Here for the reference of the reader and because the algebraic steps are fairly numerous, we derive the entropy and Bernoulli equations in the boundary layer and small-angle approximations. The following identities prove useful throughout:

\begin{equation}
U_{\nu}\nabla_{\mu}U^{\nu} = 0, \quad \Pi^{\nu}_{\,\,\mu}U^{\mu} = 0, \quad \partial_{\mu}\left(\Gamma\right) = v_{\rm r}\partial_{\mu}\left(\Gamma v_{\rm r}\right). \label{identities}
\end{equation}
The last of these is valid in the limit that $v_{\theta} \ll 1$, while the first two follow generally from the conservation of the norm of the four-velocity $U_{\nu}U^{\nu} = -1$. We write the fluid equations from Equation \eqref{radhydro} as

\begin{equation}
\rho'U^{\mu}\nabla_{\mu}U^{\nu}+\nabla_{\mu}\left[4p'U^{\mu}U^{\nu}\right]+g^{\mu\nu}\nabla_{\mu}p' = \nabla_{\mu}R^{\mu\nu}_{\rm vis}, \label{radhydro2}
\end{equation}
where

\begin{equation}
R^{\mu\nu}_{\rm vis} = \frac{8}{9}\frac{p'}{\rho'\kappa}\left(\Pi^{\mu\sigma}\nabla_{\sigma}U^{\nu}+\Pi^{\nu\sigma}\nabla_{\sigma}U^{\mu}\right).
\end{equation}
Note that we are dropping the pressure gradient (conduction) term from this expression, as it is always small in the boundary layer approximation where $\partial p'/\partial \theta = 0$ to leading order in the boundary layer thickness.

To derive the entropy equation, we take the contraction of Equation \eqref{radhydro2} with $U_{\nu}$ and use the above-stated identities to turn the left-hand side into

\begin{equation}
U_{\nu}\left(\rho'U^{\mu}\nabla_{\mu}U^{\nu}+\nabla_{\mu}\left[4p'U^{\mu}U^{\nu}\right]+g^{\mu\nu}\nabla_{\mu}p'\right) = -3U^{\mu}\nabla_{\mu}p'-4p'\nabla_{\mu}U^{\mu} = -3p'U^{\mu}\partial_{\mu}\ln\left(\frac{p'}{(\rho')^{4/3}}\right).
\end{equation}
In the last line we used the continuity equation to write

\begin{equation}
\nabla_{\mu}U^{\mu} = -\frac{1}{\rho'}U^{\mu}\nabla_{\mu}\rho'.
\end{equation}
We can use the chain rule to write the right-hand side of the contraction of Equation \eqref{radhydro2} with $U_{\nu}$ as

\begin{equation}
U_{\nu}\nabla_{\mu}R^{\mu\nu} = U_{\nu} \nabla_{\mu}\left[\frac{8}{9}\frac{p'}{\rho'\kappa}\left(\Pi^{\mu\sigma}\nabla_{\sigma}U^{\nu}+\Pi^{\nu\sigma}\nabla_{\sigma}U^{\mu}\right)\right] = -\frac{8}{9}\frac{p'}{\rho'\kappa}\left(\Pi^{\mu\sigma}\nabla_{\sigma}U^{\nu}+\Pi^{\nu\sigma}\nabla_{\sigma}U^{\mu}\right)\nabla_{\mu}U_{\nu}.
\end{equation}
This expression will be dominated by terms that are largest in terms of the boundary layer thickness, which will appear whenever we have a derivative or Christoffel symbol that varies as $\sim 1/\theta$. The largest such terms are

\begin{equation}
\Pi^{\mu\sigma}\left(\nabla_{\sigma}U^{\nu}\right)\left(\nabla_{\mu}U_{\nu}\right) = \frac{1}{r^2}\left(-\left(\frac{\partial \Gamma}{\partial \theta}\right)^2+\left(\frac{\partial}{\partial \theta}\left[\Gamma v_{\rm r}\right]\right)^2\right) = \frac{1}{r^2}\frac{1}{\Gamma^2}\left(\frac{\partial}{\partial \theta}\left[\Gamma v_{\rm r}\right]\right)^2,
\end{equation}
where in the last line we used Equation \eqref{identities}. We also ignored terms here that are of order 1 in the boundary layer thickness compared to these terms, which implies that this expression is only valid while $1/(\Gamma\delta\theta) > 1$ or $\Gamma\delta\theta < 1$, being a statement of the causal connectedness of the jet (which is consistent with the constancy of the pressure across the boundary layer). This last step completes the derivation of the entropy equation. 

To derive the Bernoulli equation, take the contraction of Equation \eqref{radhydro2} with $\Pi^{r}_{\,\,\nu}$ and the left-hand side becomes (after using Equation \ref{identities})

\begin{equation}
\rho'U^{\mu}\nabla_{\mu}U^{r}+4p'U^{\mu}\nabla_{\mu}U^{r}+\Pi^{\mu r}\nabla_{\mu}p'.
\end{equation}
In the boundary layer approximation, the convective derivative of $U^{r}$ becomes a partial derivative, so that $\nabla_{\mu}U^{r} \simeq \partial_{\mu}U^{r}$ to leading order in the boundary layer thickness. It also follows that

\begin{equation}
\Pi^{\mu r}\nabla_{\mu}p' = \Pi^{rr}\partial_{r}p' = \left(1+\Gamma ^2v_{\rm r}^2\right)\partial_{r}p' = \Gamma^2\partial_{r}p' = \frac{\Gamma}{v_{\rm r}}U^{\mu}\partial_{\mu}p',
\end{equation}
where the last line follows from the angular (and temporal) independence of  the pressure in the boundary layer approximation. Making further use of Equation \eqref{identities} then turns the left-hand side of the contraction of Equation \eqref{radhydro2} with $\Pi^{r}_{\,\,\nu}$ into

\begin{equation}
\frac{1}{v_{\rm r}}\left(\rho'U^{\mu}\partial_{\mu}\Gamma+4p'U^{\mu}\partial_{\mu}\Gamma+\Gamma U^{\mu}\partial_{\mu}p'\right).
\end{equation}
We can now use the entropy equation to write

\begin{equation}
U^{\mu}\partial_{\mu}p' = 4U^{\mu}\partial_{\mu}p'-3U^{\mu}\partial_{\mu}p' = 4U^{\mu}\partial_{\mu}p'-\left(4\frac{p'}{\rho'}U^{\mu}\partial_{\mu}\rho'-U_{\nu}\nabla_{\mu}R^{\mu\nu}\right),
\end{equation}
and hence we have

\begin{equation}
\frac{1}{v_{\rm r}}\left(\rho'U^{\mu}\partial_{\mu}\Gamma+4p'U^{\mu}\partial_{\mu}\Gamma+\Gamma U^{\mu}\partial_{\mu}p'\right) = \frac{1}{v_{\rm r}}\left(\rho'U^{\mu}\partial_{\mu}\Gamma+4p'U^{\mu}\partial_{\mu}\Gamma+4\Gamma U^{\mu}\partial_{\mu}p'-4\Gamma\frac{p'}{\rho'}U^{\mu}\partial_{\mu}\rho'+\Gamma U_{\nu}\nabla_{\mu}R^{\mu\nu}\right).
\end{equation}
Multiplying through by $v_{\rm r}$, the Bernoulli equation then becomes

\begin{equation}
\rho'U^{\mu}\partial_{\mu}\left[\Gamma\left(1+\frac{4p'}{\rho'}\right)\right] = \left(v_{\rm r}\Pi^{r}_{\,\, \nu}-\Gamma U_{\nu}\right)\nabla_{\mu}R^{\mu\nu}.
\end{equation}
We can now use the definition of the projection tensor to expand out the right-hand side:

\begin{equation}
\left(v_{\rm r}\Pi^{r}_{\,\, \nu}-\Gamma U_{\nu}\right)\nabla_{\mu}R^{\mu\nu} = \left(v_{\rm r}^2-1\right)\Gamma U_{\nu}\nabla_{\mu}R^{\mu\nu}+v_{\rm r}\nabla_{\mu}R^{\mu r} = -\frac{1}{\Gamma}U_{\nu}\nabla_{\mu}R^{\mu\nu}+v_{\rm r}\nabla_{\mu}R^{\mu r}.
\end{equation}
The right-hand side of the entropy equation can be used to replace the first term in this expression. As is true for the radial velocity, the Christoffel symbols are always subdominant in the boundary layer thickness when taking the covariant derivative of the tensor $R^{\mu r}$, and hence

\begin{equation}
\nabla_{\mu}R^{\mu r} = \frac{1}{r^2\sin\theta}\partial_{\mu}\left[r^2\sin\theta R^{\mu r}\right] = \frac{1}{r^2\theta}\frac{\partial}{\partial\theta}\left[\frac{8}{9}\frac{p'}{\rho'\kappa}\theta\frac{1}{r^2}\frac{\partial}{\partial \theta}\left[\Gamma v_{\rm r}\right]\right],
\end{equation}
where the last term is the most dominant in the ordering with respect to the boundary layer thickness and we adopted the small angle approximation. Combining the terms then yields the Bernoulli equation as given in Equation \eqref{bernoulli}.

\section{Trans-relativistic solutions}
\label{sec:trans}
When the pressure radial power-law index $n$ and density radial power-law index $q$ satisfy $q = n-1$, the Lorentz factor of the jet is a constant, which results from the balance between the increasing inertia contained in the outflow as it expands in angle and the pressure gradient that attempts to accelerate the fluid. In this case, since the radial component of the four-velocity does not depend on radius along the axis, we can maintain all of the relativistic terms in the fluid equations and recover self-similar relations. As such, we can directly test the assertion made above, that the flow should become non-relativistic at a value of $\xi$ that is of the order $\Gamma_{\rm j}$, which corresponds to an angle that is of the order $\delta\theta$ to within a factor that depends logarithmically on $\Gamma_{\rm j}$. 

Maintaining the same parameterizations as above and letting $q = n-1$ and $m = 0$, we can show that, with all of the relativistic terms maintained, the Bernoulli and entropy equations respectively become\footnote{We note that we are still assuming a relativistic equation of state here, such that the enthalpy of the radiation field dominates the rest-mass energy of the scatterers within the flow.}

\begin{equation}
\frac{d}{d\xi}\left[f\frac{dh}{d\xi}\sqrt{1+\Gamma_{\rm j}^2v_{\rm j}^2\left(\frac{df}{d\xi}\right)^2}\right] = -\frac{4}{9}\left\{\frac{\Gamma_{\rm j}^2v_{\rm j}^2h\left(\frac{d^2f}{d\xi^2}\right)^2}{\left(1+\Gamma_{\rm j}^2v_{\rm j}^2\left(\frac{df}{d\xi}\right)^2\right)^{3/2}}+\frac{\Gamma_{\rm j}^2v_{\rm j}^2\frac{df}{d\xi}}{\sqrt{1+\Gamma_{\rm j}^2v_{\rm j}^2\left(\frac{df}{d\xi}\right)^2}}\frac{d}{d\xi}\left[h\frac{d^2f}{d\xi^2}\right]\right\}, \label{bernex}
\end{equation}

\begin{equation}
\left(\frac{n}{4}-1\right)\frac{dh}{d\xi}\frac{df}{d\xi}-f\frac{d^2h}{d\xi^2} = \frac{4}{9}\frac{\Gamma_{\rm j}^2v_{\rm j}^2h}{1+\Gamma_{\rm j}^2v_{\rm j}^2\left(\frac{df}{d\xi}\right)^2}\left(\frac{d^2f}{d\xi^2}\right)^2. \label{entex}
\end{equation}
Making some substitutions and algebraic manipulations, we can combine these two equations to yield:

\begin{equation}
\frac{df}{d\xi}\frac{d}{d\xi}\left[\Gamma_{\rm j}^2v_{\rm j}^2\left(f\frac{df}{d\xi}\frac{dh}{d\xi}+\frac{4}{9}h\frac{d^2f}{d\xi^2}\right)+\frac{n}{4}h\right] = 0.
\end{equation}
Provided that $df/d\xi \neq 0$ (which is satisfied near the axis of the jet where $df/d\xi \simeq 1$), then the second derivative in this expression must be zero, which yields the following integral:

\begin{equation}
\Gamma_{\rm j}^2v_{\rm j}^2\left(f\frac{df}{d\xi}\frac{dh}{d\xi}+\frac{4}{9}h\frac{d^2f}{d\xi^2}\right) = -\frac{n}{4}h. \label{ssex3}
\end{equation}
Here the constant of integration was set to zero to be consistent with the boundary conditions along the axis, that $f(0) = h(0) = 0$ and $df/d\xi(0) = 1$. We can numerically integrate Equations \eqref{ssex3} and \eqref{entex} with the boundary conditions at $\xi = 0$ until $df/d\xi = 0$; we then define $\xi_{\rm nr}$ -- the location at which the flow becomes non-relativistic and $\Gamma \equiv 1$ -- as the point where $df/d\xi = 0$.

\begin{figure} 
   \centering
   \includegraphics[width=0.495\textwidth]{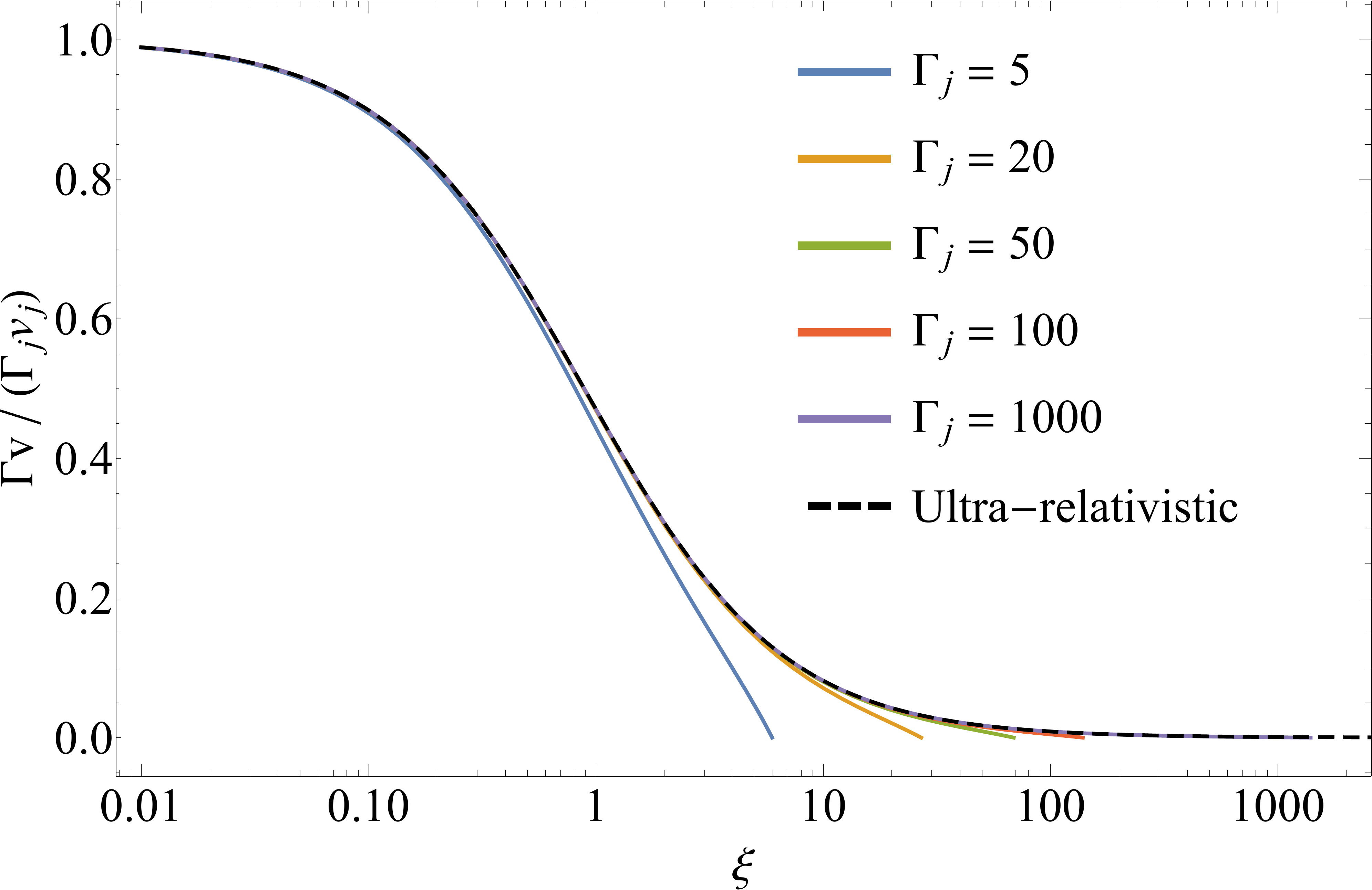} 
   \includegraphics[width=0.495\textwidth]{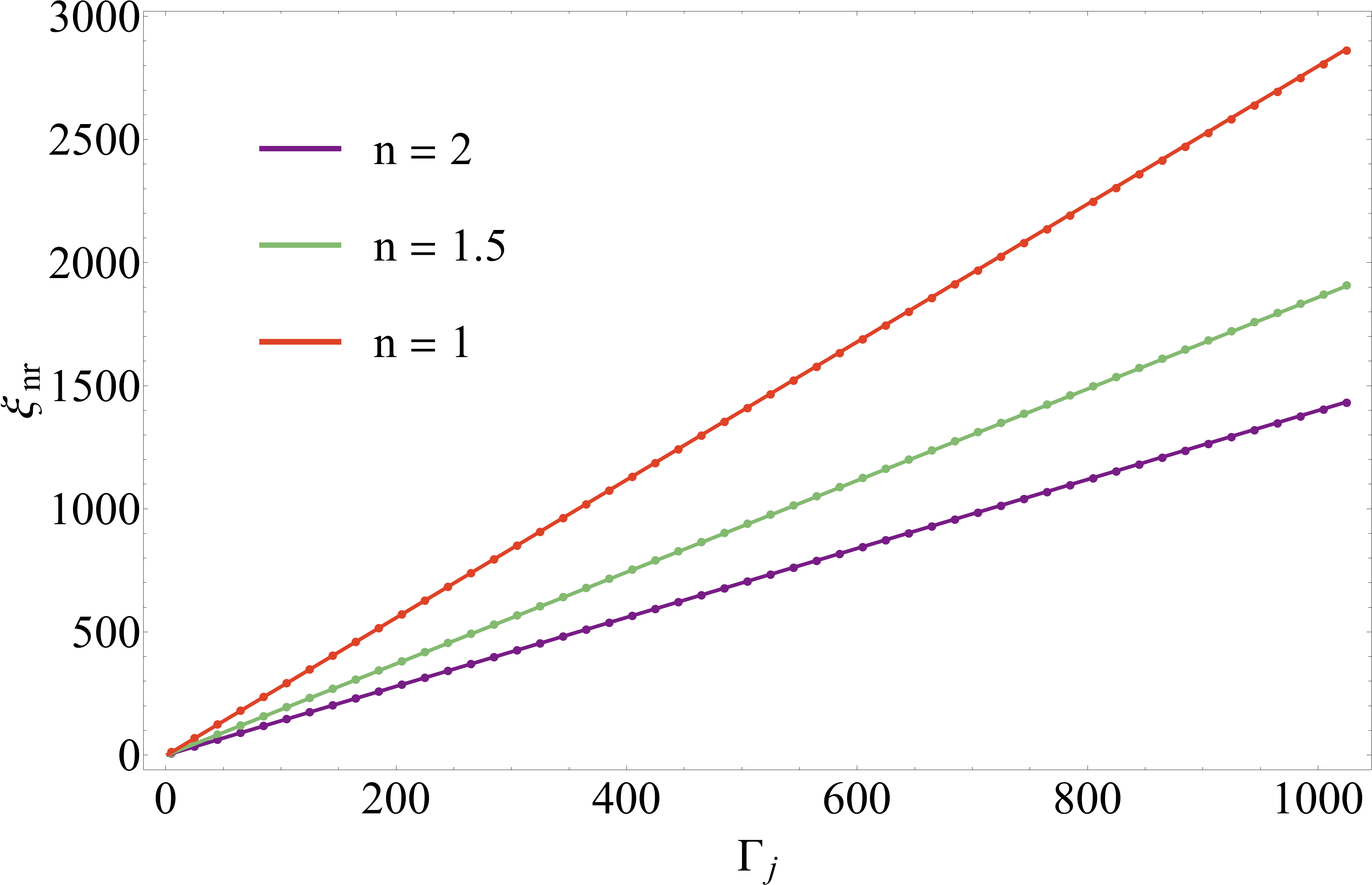} 
   \caption{Left: The radial component of the four velocity for the trans-relativistic solution with the Lorentz factors in the legend when $n = 2$ (i.e., the comoving pressure falls off as $p' \propto r^{-2}$); the black, dashed curve is the self-similar solution (Equation \ref{exsols} with $\beta = 1/2$). As the Lorentz factor of the jet becomes larger, the agreement between the trans-relativistic solution and the self-similar solution is better over a larger range of $\xi$ before the trans-relativistic solution equals zero at a finite $\xi \equiv \xi_{\rm nr}$. Right: The solution for $\xi_{\rm nr}$ (where the velocity of the trans-relativistic solution equals zero) as a function of teh Lorentz factor of the outflow. The points give the numerical values resulting from the solutions, while the curves are linear functions of $\Gamma_{\rm j}$, showing that the location at which self-similarity breaks down increases almost exactly linearly with the jet Lorentz factor.}
   \label{fig:trans}
\end{figure}

The left panel of Figure \ref{fig:trans} illustrates the radial component of the four-velocity, normalized by $\Gamma_{\rm j}v_{\rm j}$, for the Lorentz factors shown in the legend; here we let $n = 2$, so that the pressure falls off as $p' \propto r^{-2}$, and the density declines as $\rho' \propto r^{-1}$ (i.e., such that the Lorentz factor is a constant along the jet axis). The black, dashed curve shows the ultra-relativistic solution (i.e., the solution for $df/d\xi$ from Equations \ref{sstot1} and \ref{sstot2}) for comparison. We see that as the flow becomes more highly relativistic and $\Gamma_{\rm j}$ increases, the exact, trans-relativistic solution matches the ultra-relativistic one over a broader range of $\xi$. Eventually the trans-relativistic solutions equal zero velocity at a finite value of $\xi$, which we define as $\xi_{\rm nr}$. The right-hand panel shows the value of $\xi_{\rm nr}$ for the three different pressure power-law indices $n$ as a function of $\Gamma_{\rm j}$. The points give the numerical values, while the solid lines are fits of the form $\xi_{\rm nr} = C_{\rm n}\left(\Gamma_{\rm j}-1\right)$ with $C_{2} = 1.4$, $C_{1.5} = 1.86$, and $C_1 = 2.8$ (the subscript denotes the value of the pressure power-law index). These coefficients are a factor of $\sim 1.5$ larger than the analytic prediction given by Equation \eqref{xinr}, but the linear behavior between $\xi_{\rm nr}$ and $\Gamma_{\rm j}$ expected from the heuristic arguments in Section \ref{sec:energy} is apparent from this figure.

At the location where the flow velocity equals zero, the density also diverges (i.e., $dh/d\xi(\xi_{\rm nr}) = 0$), and as such the solution terminates in a ``wall'' that contains most of the mass. This feature arises from the fact that these solutions are in the presence of a pressure gradient, but nonetheless the velocity equals zero at a finite location; the only way to satisfy this condition -- that the velocity be zero in spite of the accelerating influence of the pressure gradient (absent a gravitational field) -- is to have an infinite density. More realistically, we expect near this location that the contribution of the rest-mass energy to the Bernoulli parameter will start to dominate, and the solution will either transition to a constant velocity (the coasting regime) or to a hydrostatic atmosphere in the presence of a gravitational field.

\bsp	
\label{lastpage}
\end{document}